# InAs nanocrystals with robust p-type doping

*Lior Asor, Jing Liu, Yonatan Ossia, Durgesh C. Tripathi, Nir Tessler, Anatoly I. Frenkel, Uri Banin\**

L. Asor, Y. Ossia, Prof. U. Banin
The Institute of Chemistry and The Center for Nanoscience and Nanotechnology,
The Hebrew University of Jerusalem,
Jerusalem 91904, Israel
E-mail: uri.banin@mail.huji.ac.il

Prof. J. Liu
Physics Department,
Manhattan College,
Riverdale, New York 10471, USA

Prof. A. I. Frenkel
Department of Material Science and Chemical Engineering,
Stony Brook University,
Stony Brook, New York 11794, USA

Dr. D. C. Tripathi, Prof. N. Tessler
The Zisapel Nano-Electronics Center,
Department of Electrical Engineering,
Technion – Israel Institute of Technology, Haifa 32000, Israel

ABSTRACT

Robust control over the carrier type is fundamental for the fabrication of nanocrystal-based optoelectronic devices, such as the p-n homojunction, but effective incorporation of impurities in semiconductor nanocrystals and its characterization is highly challenging due to their small size. Herein, InAs nanocrystals, post-synthetically doped with Cd, serve as a model system for successful p-type doping of originally n-type InAs nanocrystals, as demonstrated in field effect transistors (FETs). Advanced structural analysis, using atomic resolution electron microscopy and synchrotron X-ray absorption fine structure spectroscopy reveal that Cd impurities reside near and on the nanocrystal surface acting as substitutional p-dopants replacing Indium. Commensurately, Cd-doped InAs FETs exhibited remarkable stability of their hole conduction, mobility, and hysteretic behavior over time when exposed to air, while intrinsic InAs NCs FETs were easily oxidized and their performance quickly declined. Therefore, Cd plays a dual role acting as a p-type dopant, and also protects the nanocrystals from oxidation, as evidenced directly by Xray photoelectron spectroscopy measurements of air exposed samples of intrinsic and Cd doped InAs NCs films. This study demonstrates robust p-type doping of InAs nanocrystals, setting the stage for implementation of such doped nanocrystal systems in printed electronic devices.

## 1. Introduction

In the general field of "printed electronics", while great progress has been made in terms of metallic inks, there is a gap in the availability of robust inorganic semiconducting inks. Semiconductor nanocrystals (SC NCs) represent outstanding candidates for such inks for the fabrication of opto-electronic devices such as field effect transistors (FETs),[1–5] light emitting diodes,[6–10] solar cells,[11–13] and photodetectors.[14,15] This is due to the broad tunability of their electronic and optical properties via the quantum confinement effect enabled by impressive synthetic control of the NC size, composition and shape.[16,17] Additionally, the wet chemical processing of the NCs, facilitated by their flexible surface chemistry, allows for their realization in bottom-up fabrication of printed electronics devices.[18] Sophisticated use of ligands to modify[19] or polarize[20] the nanocrystals has been developed and is an important part of the toolbox for such implementations. However, doping, widely practiced in bulk semiconductors to control electrical functionality, still presents an open challenge for SC NCs both from the viewpoint of fundamental understanding as well as for device applications.[21–24] The complexities in doping SC NCs are amplified by the possible differences in behavior of the isolated individual NCs, versus their characteristics in a working NC array in an electronically active device, where modifications of the surface chemistry are essential for effective interconnection to enable electrical transport among the nanocrystals.[25–27] Such surface modifications often affect the doping state and even the nature of doping on the level of the NC array.[26,28–32] Indeed, "remote doping", through binding of electron donating groups, is a successful approach to electronically dope the NCs.[1,33–36] On the other hand, this very same process often limits the ability to generate stable n- or –p-type NCs with controlled doping levels and tailored surface chemistry in the array, thus calling for utilization of impurity doping. Availability of doped NCs will also enable bottom-up "printed electronics" with doped NCs as active semiconductor inks to form all nanocrystal-based p-n junctions,[37] that are the basis of numerous (opto)electronic devices.

To this end, we reported on impurity doping of InAs nanocrystals, which brings them to the regime of heavy doping as reflected in the optical properties, scanning tunneling spectroscopy data for single isolated nanocrystals and theoretical analysis.[21] Impurity doping was also reported for CdSe[38–40] and Pb chalcogenide[30,32,37,41–43] NCs, which serve as outstanding model systems but are limited in terms of applicability by health and environmental hazard regulations. This provides further motivation to pursuit doping of III-V SC NCs as promising building blocks for optoelectronic NC devices[29,44] due to their relatively low toxicity, high carrier mobility and their optical properties, which are tunable across the visible to the near IR. A main drawback of these materials is their susceptibility for oxidation, related to their covalent nature and a high density of surface traps. The high sensitivity of the electrical properties of InAs to adsorbed gaseous molecules causes sharp changes in conductivity, and considering the large surface area to volume ratio of nanomaterials, InAs nanowires have been studied as gas detectors.[45,46] In contrast, the need for stable performance of electronic devices in ambient atmosphere is obvious, necessitating robust control on the majority carrier type and concentration in the doped NCs.

Recently, we demonstrated the effect of n-type impurity doping of InAs NCs in field effect transistors (FETs),[47] by implementing a post-synthesis doping reaction with Cu.[21,47–50] The Cu doped-InAs NCs FETs showed improved performance over the intrinsic InAs films due to electron donating impurity sub-band, in line with the n-type doping characteristics at the individual nanocrystal level.[21,49,51] However, the fabrication of p-type InAs NCs FETs is challenging as one has to overcome the intrinsic excess free electrons of as-synthesized InAs NCs.

To date, only a single work demonstrated p-type doping of colloidal InAs NCs. Geyer et al.[52] incorporated Cd impurities on the surface of InAs NCs by addition of Cd precursor to InAs NCs at late stages of the synthesis and refluxing at high temperatures. However, the exact mechanism of doping remains unclear. The authors suggested two possible mechanisms that may explain the observed change in conductivity. The first is the substitution of In by Cd, forming acceptor states near the valence band. The second is surface attachment of Cd onto the nanocrystals and passivation of intrinsic donor surface states. Thus, a thorough understanding of this phenomenon is lacking, and additional data is needed to gain insight into the mechanism of doping. Beyond the fundamental interest, addressing the InAs-dopant interaction and elucidating the doping mechanism in this system, paves the way to the use of additional less toxic, potentially p-type dopants to InAs and contributes to the understanding of impurity doping processes and outcomes in SC NCs in general.

In this work, we develop a method for *post*-synthesis doping of InAs NCs with Cd achieving robust p-type characteristics in FETs with enhanced stability in ambient conditions. The post-synthesis doping approach allows to separately control the NC size and later the impurity levels, and also enables the direct comparison of the NC characteristics before and after doping. To decipher the doping mechanism, we utilize structural analysis, including high resolution scanning transmission electron microscopy (HR-STEM), X-ray diffraction (XRD) and X-ray absorption fine structure spectroscopy (XAFS) revealing that Cd atoms act as substitutional dopants replacing Indium near the surface of the NCs, and leading to the observed p-type doping. This doping configuration also has positive consequences on the sensitivity of the FETs to the ambient atmosphere. While undoped InAs NCs FETs were easily oxidized and their performance declined when exposed to air, the Cd-doped FETs exhibited much improved stability over time as also reflected in their robust hysteretic behavior. Therefore, the surface location of the Cd in the NCs, not only acts as a p-type dopant, but also as a protective layer, preventing oxidation of In and As, as evidenced directly by Xray photoelectron spectroscopy (XPS) measurements. This study sets the stage for future fabrication of InAs NCs based p-n homojunction and additional more sophisticated devices, bringing closer the realization of SC NC inks with controlled doping for printed electronics applications.

## 2. Results and Discussion
### 2.1. Doping reaction and structural characterization

InAs NCs were synthesized using the well-established method employing $InCl_3$ and tris(trimethylsylilarsine) as precursors in triotylphosphine (TOP) followed by size-selective separation to obtain a large fraction of NCs with a diameter of 5nm.[53] Inspired by a method for shell growth,[54] as illustrated in **Figure 1**a, post-synthesis doping with Cd was achieved by dropwise addition of $Cd(Oleate)_2$ for achieving 3 different nominal In:Cd ratios of 1.0:0.35, 1.0:0.7, and 1.0:1.1, over 20 minutes, to a preheated solution of InAs NCs in 1-octadecene and oleylamine at 260°c under inert atmosphere, and the reaction was left for 3 hours. Focusing on the sample with the highest Cd content, with time after Cd addition, the photoluminescence (PL) increases sharply while the absorption feature is red shifted and broadens (Figure S1). The absorption and PL spectra of the original InAs NCs and of the purified Cd-doped InAs NCs sample are presented in Figure 1b, c. The first exciton absorption peak is broadened and red shifted by 95 meV after Cd incorporation. Correspondingly, the PL peak is also red shifted, but to a larger extent of 190 meV, and notably, its intensity increases by an order of magnitude relative to the undoped InAs NCs sample.

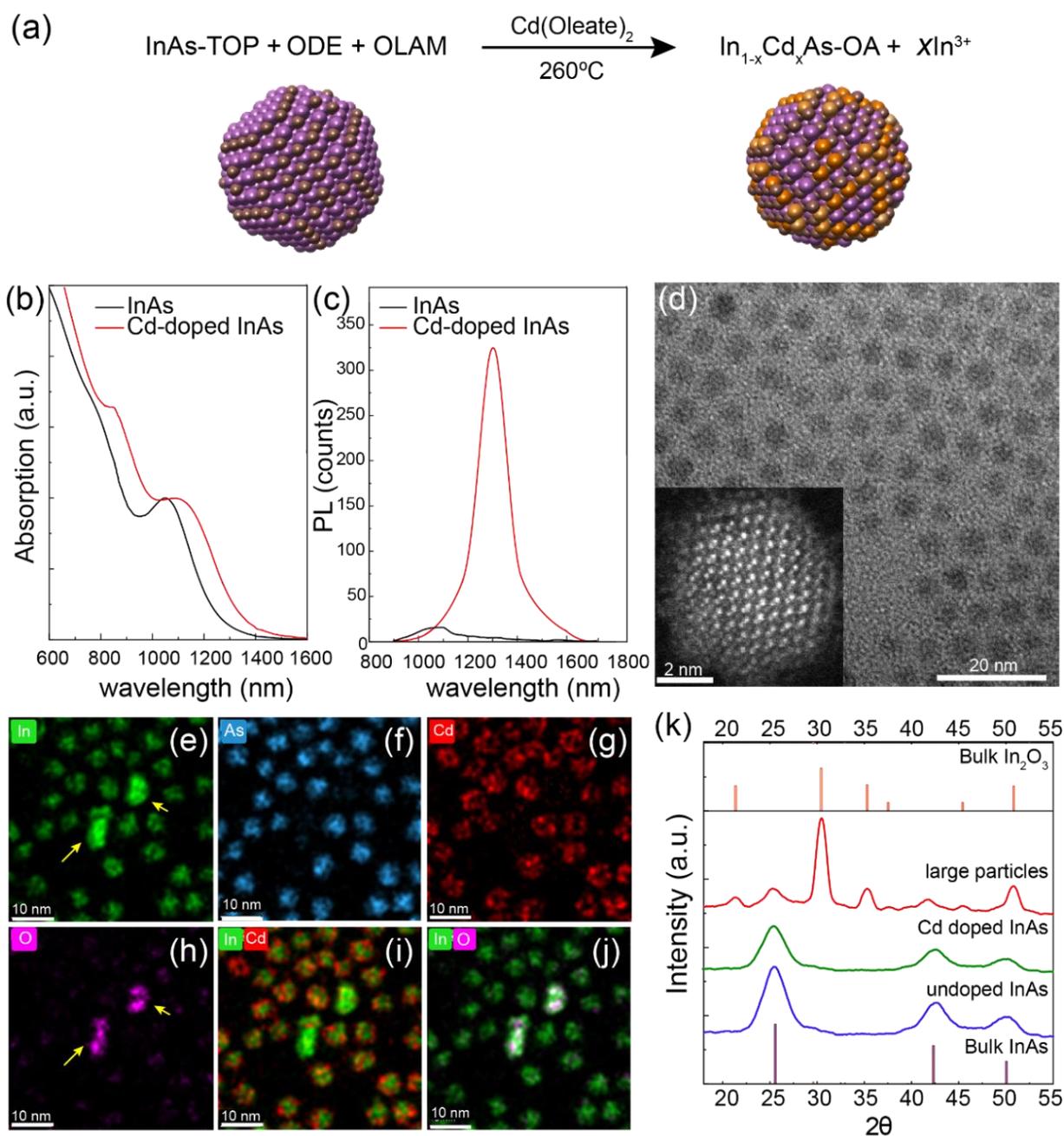

**Figure 1.** (a) NC doping reaction scheme. (b-c) Absorption and photoluminescence of as synthesized (black lines) and Cd doped (red lines) InAs NCs. (d) TEM image of the Cd doped InAs NCs. Inset: high resolution HAADF STEM image of a single NC projected from the <110> zone axis. (e-j) EDS-STEM images of the Cd doped InAs QDs with elemental mapping showing In (green), As (pale blue), Cd (red), O (purple), In-Cd overlay, and In-O overlay. Remaining large $In_2O_3$ NCs are identified. (k) X-ray diffraction pattern of zincblende, as-synthesized (blue) and purified Cd doped InAs (green) NCs, together with the fraction containing large particles (red) showing the X-ray diffraction pattern of $In_2O_3$ phase. Bulk InAs and $In_2O_3$ XRD peaks are shown on the bottom and top, respectively.

Transmission electron microscopy (TEM) characterization comparing the original InAs NC fraction and the product of the doping reaction show no significant apparent change in the diameter of the NCs, but a small amount of large particles (20-30nm in diameter) is also formed during the doping reaction (Figure S2). Using size selective precipitation method, we successfully separated out most of the large particles resulting in a pure fraction of Cd-doped InAs NCs as the TEM image shows in Figure 1d. Inductively-coupled plasma atomic emission spectroscopy (ICP-AES) characterization of the original InAs NCs yields an In:As ratio of 1.2:1, while the purified Cd-doped sample shows an In:As:Cd ratio of 1.07:1:0.5 (Table S1). The reduction in the In:As ratio already implies some substitution of In by Cd upon doping.

We further characterized the purified Cd-doped InAs NCs samples by HR-STEM. Figure 1 (e-j) shows high resolution energy dispersive X-ray spectroscopy (EDS) mapping, identifying In, As, Cd and O elements within the sample. The remaining larger particles in the purified sample indicated by the arrows lack the arsenic and Cd EDS signals, but Indium and Oxygen are strongly observed. Powder XRD measurements confirm the existence of an $In_2O_3$ phase in the fraction that contains these large particles (Figure 1k, red trace). This side product was reported also in the incorporation of Cd and Zn on the surface of InP NCs, where the proposed formation mechanism was by the high temperature reaction of freed In being substituted by either Cd or Zn on the surface of the NCs, reacting with carboxylate molecules that are present in the reaction flask thus forming $In_2O_3$ particles.[55] Indeed, reducing the amount of free oleic acid in the Cd(Oleate)$_2$ precursor solution by using a stoichiometric ratio in the precursor preparation resulted in a reduced amount of $In_2O_3$ particles as a side product. For the main doped NC fraction, the XRD peaks conform well with the bulk InAs zincblende pattern and the spectrum is hardly changed with respect to the original NCs. The EDS maps of the doped NCs show that In and As are spread evenly, while Cd is found predominantly on the surface as is emphasized by the overlay image of In and Cd signals (Figure 1i).

For further clear indication of doping, and for uncovering its local structural details from the dopant (Cd) and host (In and As) atom perspectives, we used X-ray absorption fine structure (XAFS) spectroscopy that will be described later. With our goal being the achievement of robust p-type doping, we next describe the results from electrical measurements in FETs for these particles.

## 2.2. Field Effect Transistors - basic electrical characterization

InAs NCs thin film transistors were prepared following our optimized approach reported recently.[47] Briefly, a solution of InAs NCs was spin cast on top of heavily doped Silicon substrates covered with 100 nm thick $SiO_2$ and a 10 nm thick $HfO_2$ layer that was deposited on top (see supporting information for details). In order to reduce the distance between the NCs and improve the conductivity of the films, we performed a standard solid-state ligand exchange with 1,2 Ethanedithiol (EDT), a short bi-linker ligand that is also typically doping neutral and commonly used in quantum dot solids. The ligand exchanged NCs films were homogeneous and densely packed with an average thickness of 35 nm. Prior to Source/Drain Au electrodes evaporation, thermal annealing of the NCs films at 220°C for 30min was performed under vacuum or nitrogen with the latter providing the best electrical response. During this step, organic residues desorb from the films surface and the conductivity of the films further improves, while the NCs retain their size and shape without fusion (Figure S3).

Electrical characterization of the FETs was performed under inert atmosphere. **Figure 2** shows the output and transfer characteristics of undoped and Cd-doped InAs NCs FETs. Conforming with our earlier report,[47] output characteristics of as-synthesized intrinsic InAs NCs FETs exhibit n-type conduction with clear linear and saturation regimes of operation (Figure 2a). The device is highly conductive with the drain current reaching over 1 mA at gate-source voltage of 30 V with leakage current reaching a few hundreds of nA (Figure S4). The transfer curve of the undoped InAs NCs (Figure 2b) indicates a clear n-channel with high electron mobility of 0.31 cm$^2$V$^{-1}$s$^{-1}$ and small hysteresis.

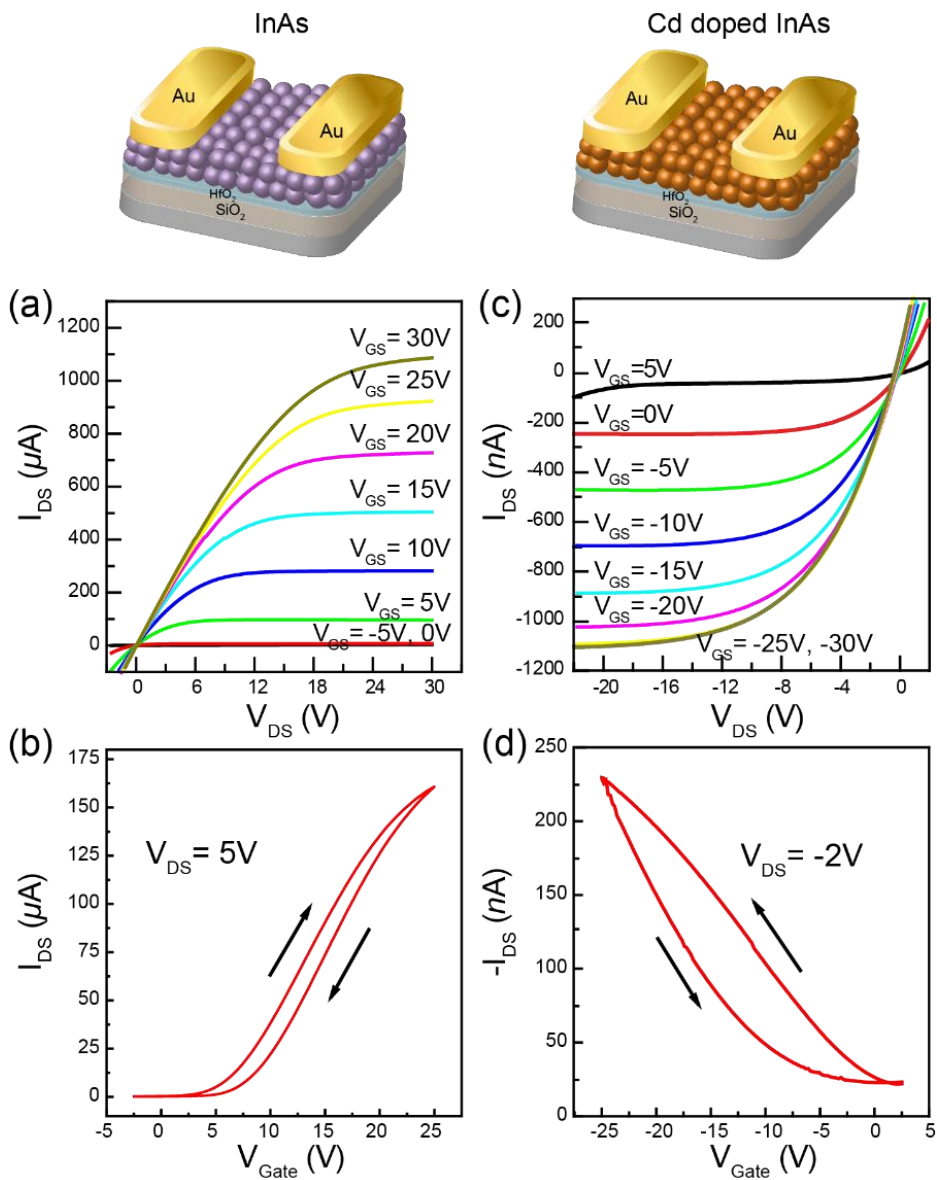

**Figure 2.** Electrical characterization of intrinsic and Cd-doped InAs NCs FETs. (a-b) Output and transfer curves, respectively, of intrinsic InAs NCs showing n-type conduction. (c-d) Output and transfer curves, respectively, of Cd doped InAs NCs showing successful doping and p-type conduction.

Cd-doped FETs presented fundamentally different characteristics compared to the devices of undoped InAs NCs. A unipolar p-type conduction channel is seen, suggesting switching of the majority carrier type upon Cd doping. The output characteristics also show distinguishable linear and saturation regimes (Figure 2c) with low leakage current (Figure S4). Cd treatment for shorter times (i.e. less than 3 hours under 260∘c) resulted in n-type conduction even though the Cd concentration on the surface of the NCs is similar to that detected for the p-type Cd-doped InAs NCs. This further indicated that surface passivation alone is not sufficient for doping the NCs, but rather substitutional doping is required for the switching of the majority carrier from electrons to holes. Devices prepared with lower Cd content manifested ambipolar characteristics as depicted in the transfer characteristics of these devices shown in Figure S5. The ability to switch the carrier type with respect to the gate-source voltage is presented, as the hole and electron conduction threshold voltages equal -11V and +18V, respectively.

The transfer characteristics of the P-type Cd doped InAs FET (Figure 2d) show lower conductivity of holes compared to the n-type device and the linear hole mobility is $1.5 \cdot 10^{-3}$ cm$^2$V$^{-1}$s$^{-1}$, more than two orders of magnitude lower compared to the electron mobility of intrinsic InAs NC FETs. High degree of hysteresis is also observed. Holes in bulk InAs are 20 times heavier than electrons and the hole mobility is reported to be lower by two orders of magnitude.[56] Moreover, the large hysteresis indicates that holes are trapped more easily in the doped InAs NC film thus also lowering mobility, as will be further analyzed below. Nonetheless, the control on the carrier type in InAs is achieved and the p-type nature of the Cd-doped InAs nanocrystals was demonstrated. There are several possibilities that can explain the p-type doping in this case, in particular considering the initial n-type character of the InAs NC films, as was discussed in the early work of Bawendi and co-workers[52]: Firstly, cadmium serving as a substitutional impurity replacing indium. Due to $Cd^{2+}$ oxidation state, compared with the $In^{3+}$, such substitutional doping will result in generation of free holes. At sufficient concentration, these will compensate for the electron concentration in the original n-type InAs NCs and generate the hole carriers. Second, cadmium on the NC surface passivates the InAs NC donor states while creating an acceptor state, leading to the observed p-type character. The EDS-STEM analysis discussed above already indicated high abundance of Cd on the NC surface and in order to further scrutinize the precise doping mechanism, detailed local structural analysis of dopant and host atoms is needed, as described in the next section.

**2.3. XAFS study for the analysis of the origin of the doping**

To gain a better understanding of the observed p-type doping mechanism, X-ray absorption fine structure (XAFS) spectra of In, As and Cd were collected for intrinsic and Cd doped InAs NC samples at the QAS beamline of National Synchrotron Light Source - II (see SI for experimental details). Focusing on the sample with the highest Cd content, both indium and arsenic K-edge spectra remain unchanged upon Cd doping in all the samples, indicating that the local InAs structure is largely maintained even for the highest nominal doping levels (SI Figure S6). For Cd located as either a bulk or surface substitutional impurity, the indium K-edge XAFS spectrum is not expected to be very sensitive to the doping reaction as In is not in the nearest coordination shell of Cd.  Additionally, by taking into account the similar size of Cd and In atoms and lack of their spectral contrast due to their neighboring positions in the Periodic Table, the local environments of arsenic atoms (and thus As K-edge XAFS spectra) are also expected to remain the same after doping. We note that a similar effect  - a relatively low change in the In and As K-edge XAFS data with Ag doping -  indicated, too, a substitutional (as opposed to interstitial) doping location at the QD surface.[48]

Investigating the Cd XAS data provides the complementary information for the nature of doping (**Figure 3**). K-edge X - ray absorption near edge structure (XANES) data of both Cd(Oleate)$_2$ and the Cd-doped InAs NCs powdered samples with different doping levels are not shifted in energy. This indicates that Cd retains its 2+ oxidation state during the doping reaction (Figure 3a), consistent with its possible p-type substitutional doping in InAs but it is also the expected valency of surface Cd bound to oleate. This is further clarified upon inspecting the Fourier transform of the Cd K-edge extended XAFS (EXAFS) data of the Cd-doped sample (Figure 3b, black line), which shows two dominant scattering paths implying presence of two different species in very close proximity to Cd. A model describing Cd solely as a substitutional dopant (for In) in the bulk of the NCs failed to reproduce the acquired data, since such scenario would lead to a single scattering path (Cd-As) in the EXAFS. Considering the HR-STEM results shown earlier, indicating that Cd is predominantly found on the surface of the NCs, we hypothesize that given that the intrinsic InAs NCs surface is indium rich (as indicated by ICP, table S1), Cd substitutes In atoms on the surface of the NCs and will thus interact with the nearest neighboring As atoms. In this scenario, Cd is the cation that is exposed to the surface of the NCs and will interact also with the surface ligands, most likely the oleate molecules originating from the Cd precursor. Therefore, an additional interaction of Cd with oxygen from the oleate molecules is expected in the Cd K-edge EXAFS data. The hypothesis is well supported by the EXAFS data. The Cd edge EXAFS data of a Cd(Oleate)$_2$ standard sample (Figure 3b, blue dashed line) shows a peak for Cd-O interaction at ~1.8Å, correlating well with the first peak of the Cd-doped InAs sample. In addition, given the similar size of Cd and In, the In-As interaction peak at 2.4Å (Figure 3b, red dotted line), correlates well with the second peak of the Cd edge consistent with a Cd-As interaction as well.

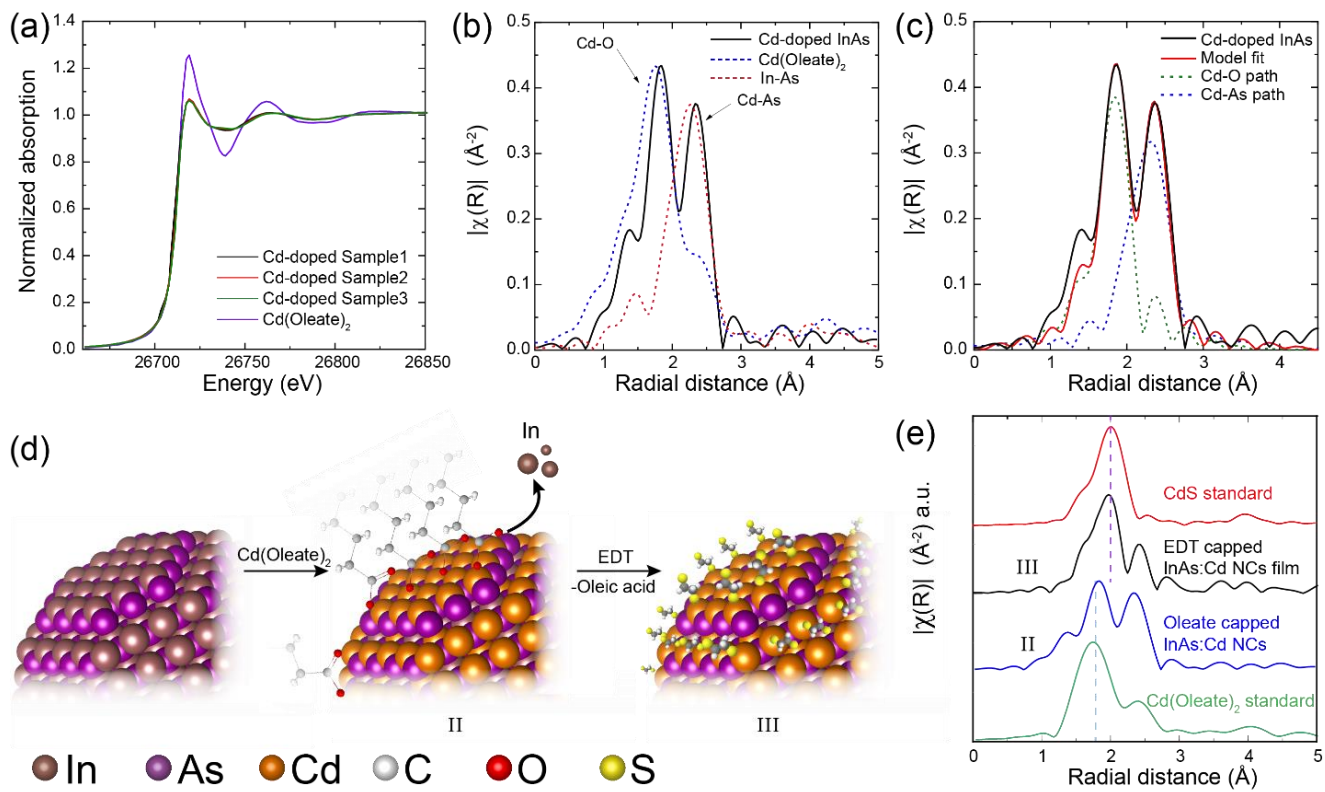

**Figure 3.** (a) Normalized Cd K-edge XANES spectra of Cd-doped InAs NCs samples and Cd(Oleate)$_2$ standard sample. (b) Fourier transform magnitude of k-weighted EXAFS data of Cd-doped InAs NCs sample with the highest Cd content (k-range: 2-11 Å$^{-1}$; black squares), Cd(Oleate)$_2$ precursor sample (k-range: 2-11 Å$^{-1}$; blue dots) and In k-weighted $\chi(R)$ EXAFS of InAs NCs sample (k-range: 2-11 Å$^{-1}$; red dots). (c) Data and fit of the first coordination shell contributions to Cd K-edge EXAFS data of the same Cd-doped InAs NC sample using the structural model of a Cd impurity occupying surface substitutional site is shown together with the calculated Cd-O (dashed, green) and Cd-As (dashed, blue) scattering paths. (d) Illustration of the above model where Cd substitutes In on the surface of InAs NC. (e) Fourier transform magnitude of k-weighted Cd edge EXAFS data (k-range: 2-11 Å$^{-1}$) for as synthesized Cd doped InAs NCs capped with Oleate and Ethanedithiol capped Cd doped InAs NCs films. Cd(Oleate)$_2$ and CdS standards are also shown for comparison.

In line with the hypothesis, we devised a model that combines Cd-As interaction (corresponding to Cd placement in InAs lattice as a substitutional dopant) with Cd-O interaction in Cd-oleate (corresponding to the surface Cd) to fit the Cd K-edge EXAFS data. This model fits the data well (Figure 3c, red line). It is observed also that the Cd-O and Cd-As paths are aligned well with the respective peaks of the acquired data, explaining the observed double-peak structure in the experimental data described above.

The coordination numbers extracted from the fits are 1.4 for Cd with As, and 1.2 for Cd-O (Table S2 for the detailed fitting parameters). Due to the ensemble nature of the XAFS technique, the presence of two contributions (Cd-O and Cd-As) in the data could represent also two populations for Cd, one- substitutional in bulk and the other - on the surface, but the correlation with the EDS-STEM analysis implies dominance for Cd on the NC surface. Cd may bind to As rich surface facets but these are not abundant in the original NCs that has excess In, as evidenced in the ICP and EDS characterizations. The most plausible explanation emerging from the structural analysis, as illustrated in Figure 3d, is therefore of Cd substituting for In on the surface. Surface cadmium is also bound and stabilized by the oleate ligand, in line with the observation of the Cd-O interactions. Similar observations were reported for Cd- and Zn-doped InP NCs.[55,57]

To verify this, and to mimic as best as possible the Cd-doped InAs NCs FETs, we performed XAFS measurements directly on films of doped NCs that underwent a solid-state ligand exchange with 1,2 Ethanedithiol (EDT), and compared the data with that of as-prepared Cd-doped InAs NCs, capped with the native oleate ligands (Figure 3e). The Cd-O interaction peak of the as-prepared Cd-doped InAs sample is diminished after the solid state ligand exchange, while the EDT-exchanged Cd-doped InAs NCs sample reveals the emergence of a new peak at a radial distance closer to 2 Å. EDT is bound to the Cd rich surface of the doped NCs by the thiol group, creating

Cd-S bonds. Cd k-edge EXAFS data of a CdS standard is well-aligned with the main peak of the EDT-capped Cd-doped sample, indicating that successful removal of oleate ligands was achieved and that the Cd-O interaction is primarily due to the interaction with the native ligands, rather than chemical oxidation on the NCs surface.

A model of Cd atoms bound to arsenic and sulfur atoms fits well to this data. From the fitting parameters we found that the Cd-As coordination number remains unchanged. The Cd-S coordination number (2.5) increases compared to Cd-O (1.2) in the Cd-doped InAs samples prior to ligand exchange, probably due to the larger size of the oleate molecule that are more sterically hindered resulting in less dense packing of oleate ligands compared to EDT.

**2.4. Atmospheric response of the FETs**

We return now to a detailed analysis of the FETs, studying the response to different atmospheric conditions crucial for demonstration of robust p-type doping in this system, and providing further understating of the doping mechanisms. To this end, we performed continuous measurements on intrinsic and Cd-doped InAs NCs FETs under different atmospheres, switching repeatedly between vacuum ($10^{-4}$ mbar base pressure) and either pure Nitrogen or dry air at different pressures. **Figure 4**a shows the drain current at $V_{gs}$ = 30V and $V_{ds}$ = 7.5V (referred to as the 'on-current') of intrinsic InAs NCs FET. The on-current decreases drastically, by an order of magnitude, upon exposure to both Nitrogen and air at different pressures. The on-current drops much faster and recovers slower when it is exposed to air, rather than pure $N_2$, indicating its added sensitivity to oxygen (Figure S7 presents representative complete transfer curves for the different atmospheric conditions). Notably, intrinsic InAs nanowire FETs, show only a 2-fold decrease in the current upon switching the atmosphere from vacuum to air.[46] The higher sensitivity of the NCs FET to the atmosphere is consistent with their much larger surface-to-volume ratio compared with the nanowires.

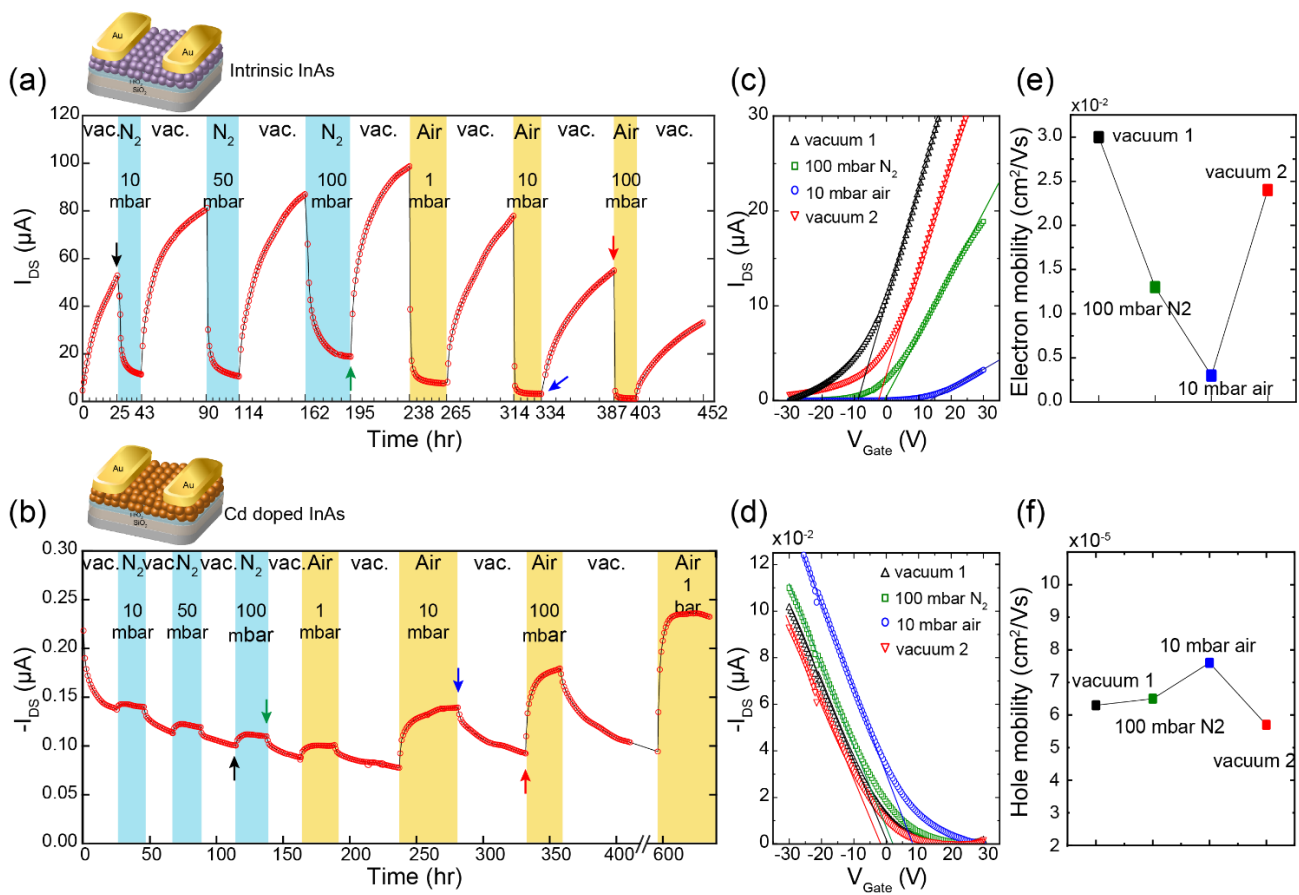

**Figure 4**. Electrical characterization of intrinsic and Cd-doped InAs NCs FETs under different atmospheres. (a, b) the drain current at $V_{ds}$= +7.5V (-7.5V) and $V_{gs}$ = +30V (-30V) of intrinsic (Cd-doped) InAs NCs FETs measured continuously, switching between vacuum (1E-4 mbar), $N_2$, and dry air, at different pressures. (c, d) Transfer curves of intrinsic and Cd doped InAs NCs FETs, respectively, measured under different atmospheres at specific stages during the continuous measurements. The time of each transfer curve presented is marked by an arrow in graphs (a) and (b). (e, f) electron and hole mobility calculated for intrinsic and Cd-doped InAs FETs under different atmospheres.

The different response to nitrogen and oxygen can be explained by different physisorption and chemisorption rates for each component on the NC surface. As proposed for PbSe NCs films that were exposed to nitrogen and oxygen[58] - nitrogen adsorbs on the NC by weak physisorption, and the adsorption is reversible at any pressure, whereas the oxygen adsorption can be described by the precursor-mediated dissociative adsorption mechanism.[59] Oxygen is first physisorbed on the NC surface, then it can either desorb back into vacuum or dissociate to react with the surface and form covalent bonds, an irreversible process. This irreversible process is pressure dependent, but even 1 mbar of dry air leads to some irreversible decrease in the drain current after vacuum due to chemisorption of oxygen.

P-type Cd-doped FETs show completely different response to changes in environment compared to the intrinsic device (Figure 4b). An opposite trend for the on-current is observed (at $V_{gs}$ = -30V and $V_{ds}$ = -7.5V) and the holes conductivity increases slightly after introducing $N_2$ and air molecules at different pressures. The Cd-doped device response is extremely weak for all nitrogen pressures, as well as for low pressures of air. The on-current for 100mbar nitrogen and 1mbar dry air transfer curves are almost unchanged compared to vacuum and the change is only up to about a factor of 2.5 even upon exposure to ambient air pressure.

$I_{ds}$ vs. $V_{gs}$ measurements show that the observed changes in conductivity in both intrinsic and Cd-doped NC devices are related to a combined effect of charge transfer between the NCs film and the adsorbed molecules, and changes in carrier mobility. Figure 4c, d present the transfer characteristics of the intrinsic and Cd-doped FETs, respectively, monitored during the time dependent measurements at specific stages marked by arrows in Figure 4a, b. For both intrinsic and Cd-doped InAs NCs FETs, the threshold voltage becomes more positive as the atmosphere changes from vacuum to air, meaning that electrons are depleted from the film surface and transferred to the adsorbed molecules. We believe that the surface electron accumulation layer and surface defect states of intrinsic InAs NCs facilitate electron transfer from the NCs surface to the adsorbed electron-withdrawing oxygen molecules forming superoxide, as suggested in other works regarding the effect of adsorbed oxygen on the performance of FETs.[45,46,58] Oxygen molecules are known to induce p-type doping in Pb chalcogenide NCs and carbon nanotubes by forming in-gap acceptor states near the valence band.[58,60,61] This leads to a decreased electron concentration and current in the intrinsic n-type device upon air exposure. The same happens for the Cd-doped, p-type InAs devices, but in this case, as electrons are depleted from the surface, the effective hole concentration consequently increases. Notably, these changes are less pronounced and we assign this to the Cd dopant layer on the surface of the NCs serving as a protective shell.

We also calculated the mobility values in each stage (Figure 4e, f). After exposure to 100 mbar of $N_2$ gas for about a day, the electron mobility of the intrinsic device drops from its initial value in vacuum. Upon exposure to 10 mbar of dry air a further significant decrease is observed, down by an order of magnitude compared with the initial mobility. Such decrease in electron mobility could be related to additional scattering paths formed as the molecules are physisorbed on the NCs surface. In contrast, the hole mobility in the p-type Cd-doped device remains intact with only minor changes, in-line with its significantly lessened sensitivity to atmospheric conditions. Actually, a small increase in hole mobility is observed as oxygen is adsorbed on the surface of the NCs film, possibly passivating defects.

**2.5. Subthreshold conduction and hysteresis response to atmospheric molecules**

While in the gated regime – one induces additional charges externally by applying gate voltages thus changing the charge carriers' density within the semiconductor layer, the subthreshold current better represents the native electronic characteristics of the device. In addition, investigation of the hysteretic behavior of the transfer curves, which is underexplored for NCs-based FETs, holds important information on the role of adsorbed molecules-induced traps and their origin. Therefore, we analyze the effect of atmospheric gasses on the subthreshold current (referred to as the "off current"), as well as the hysteretic behavior of both intrinsic and Cd-doped devices, which further emphasizes the difference between the devices and the insensitivity of the p-type InAs NCs device to nitrogen and oxygen molecules at mild pressures.

**Figure 5**a-d presents the full transfer curve sweeps of both intrinsic (a,b) and Cd-doped (c,d) InAs NCs FETs before and after exposure to either 100 mbar of nitrogen (a,c) or dry air (b,d). Under the influence of environmental changes the 'On' and 'Off' currents in the transfer curves vary differently in magnitude in the two systems (insets of Figure 5a,b). For the intrinsic device, the subthreshold current drops by more than two orders of magnitude upon nitrogen exposure, while the 'On' current is reduced by less than 1 order of magnitude. The difference between 'On' to 'Off' values increases significantly with exposure to dry air compared to nitrogen, with the subthreshold current decreasing by ~4 orders of magnitude due to exposure to dry air while the 'On' current is changed by ~1 order of magnitude (Figure 5b). Additionally, the response times for the 'Off' current are approximately half of the 'On' current response upon both nitrogen and dry air exposures, while the recovery rates for the subthreshold current under vacuum are much slower than for the 'On' current (Figure S8). The recovery rate upon vacuum application is also strongly dependent on the identity of the molecules that were adsorbed on the surface of the device, with faster recovery for nitrogen relative to air consistent with nitrogen molecules binding less strongly then oxygen to the InAs surface.

A different behavior is seen for the Cd doped FETs in Figure 5c-d. The difference between the 'On' and 'Off' current is less than 1 order of magnitude and both remain stable upon exposure to either nitrogen or dry air (insets of Figure 5c-d). This is another indication that the p-type Cd-doped InAs NCs are significantly less sensitive to oxygen compared to the intrinsic InAs NCs FETs, which are already responding significantly to even 1mbar of air exposure, further establishing the dual role of Cd as both a p-type dopant and serving as a protective layer.

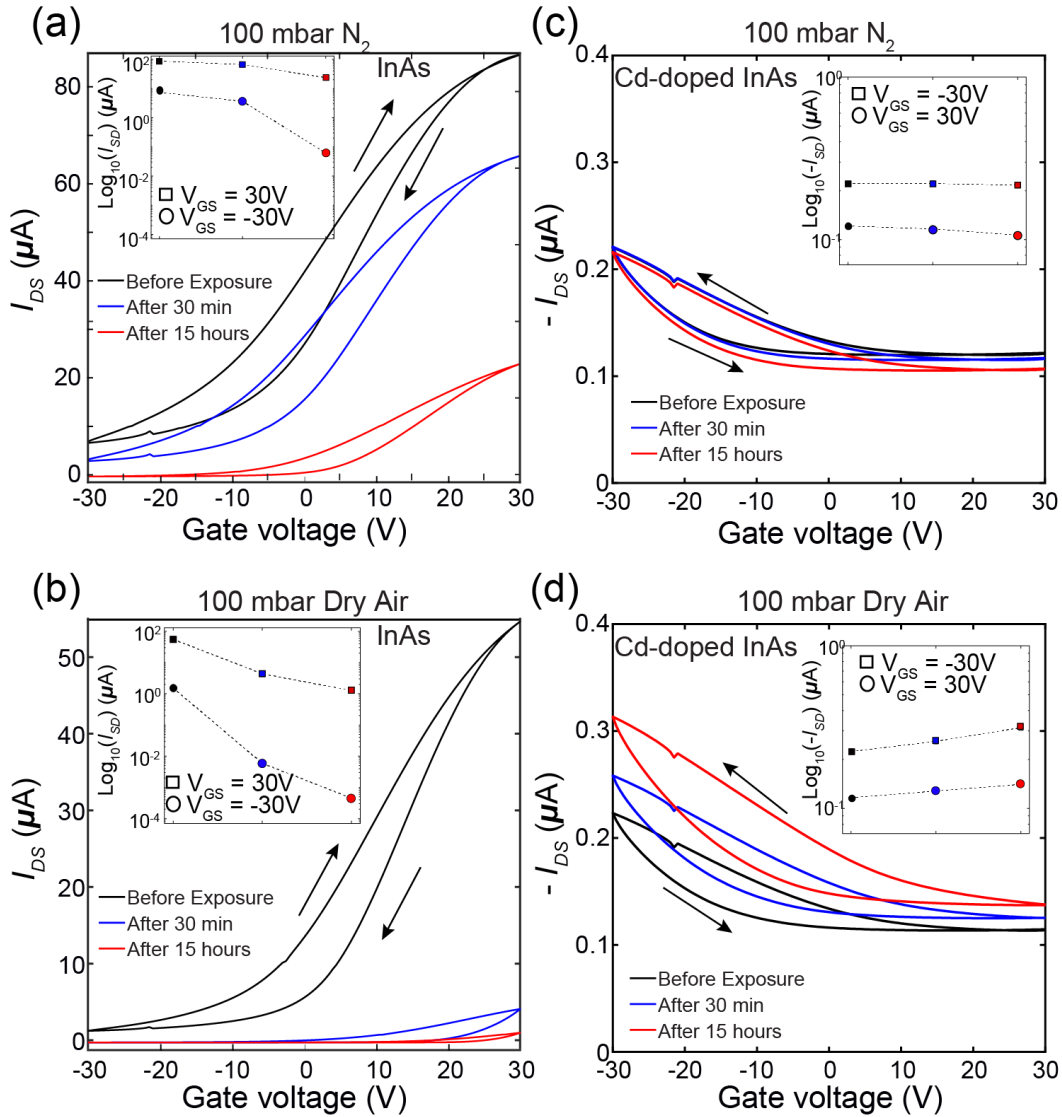

**Figure 5.** A comparison between transfer curves of as-synthesized InAs based FETs (a-b) and Cd doped InAs based FETs (c-d), at different times under exposure to 100 mbar $N_2$ and Dry Air. In each graph, the inset presents the logarithmic magnitudes of the 'On' and 'Off' currents measured at $V_{gs} = \pm 30V$ for better comparison.

The transfer curves of both intrinsic and Cd doped InAs FETs show hysteretic properties, which are related to charge carrier trapping in the NCs film. However, the large difference observed between vacuum and air exposed transfer curves makes it difficult to compare the effect of the environment on the hysteresis of the intrinsic device. Therefore, we focused on the hysteresis of the transfer curves of the Cd-doped InAs NCs FET.

**Figure 6** a,b presents the full transfer characteristics loop of Cd-doped InAs NCs FET, measured either under vacuum ($10^{-4}$ mbar) or under 100 mbar of dry air, respectively, at different sweep times. Both sets of transfer curves show lower Back-Sweep-Current hysteresis, which can originate from minority or majority charge carrier trapping. Looking closer into the transfer curves, it can be noticed that they differ in shape. When the device was measured under vacuum, the forward sweeps (from positive to negative $V_{gs}$) present a linear dependence of $I_{ds}$ versus $V_{gs}$ indicating constant hole mobility. In contrast, upon exposure to air, the forward sweeps turn nonlinear indicating hole mobility dependent on $V_{gs}$. Such differences in the transfer characteristics and hysteretic behavior are assigned to different carrier trapping mechanisms.

To better understand the origins of the observed hysteresis in the nanocrystals-based FETs we revert to the prior work on p-type Organic-field effect transistors (OFETs), in which this was ascribed to the presence of either minority (electron) or majority (holes) carrier traps.[62] In the first case there are long-lived majority carrier (hole) traps. During forward sweeping of $V_{gs}$ (from positive to negative bias), hole traps are rapidly filled, and an equilibrium between trapped holes and mobile holes density is maintained manifesting a linear transfer curve and $V_{gs}$− independent mobility. When sweeping backwards from On-to-Off, long lived hole traps are slowly emptied, leading to a decreased amount of mobile holes relative to the $V_{gs}$-dependent equilibrium density and thus the current is lower than that of the forward sweep.[62,63] This behavior is in agreement with the case of the Cd-doped NC FET under vacuum.

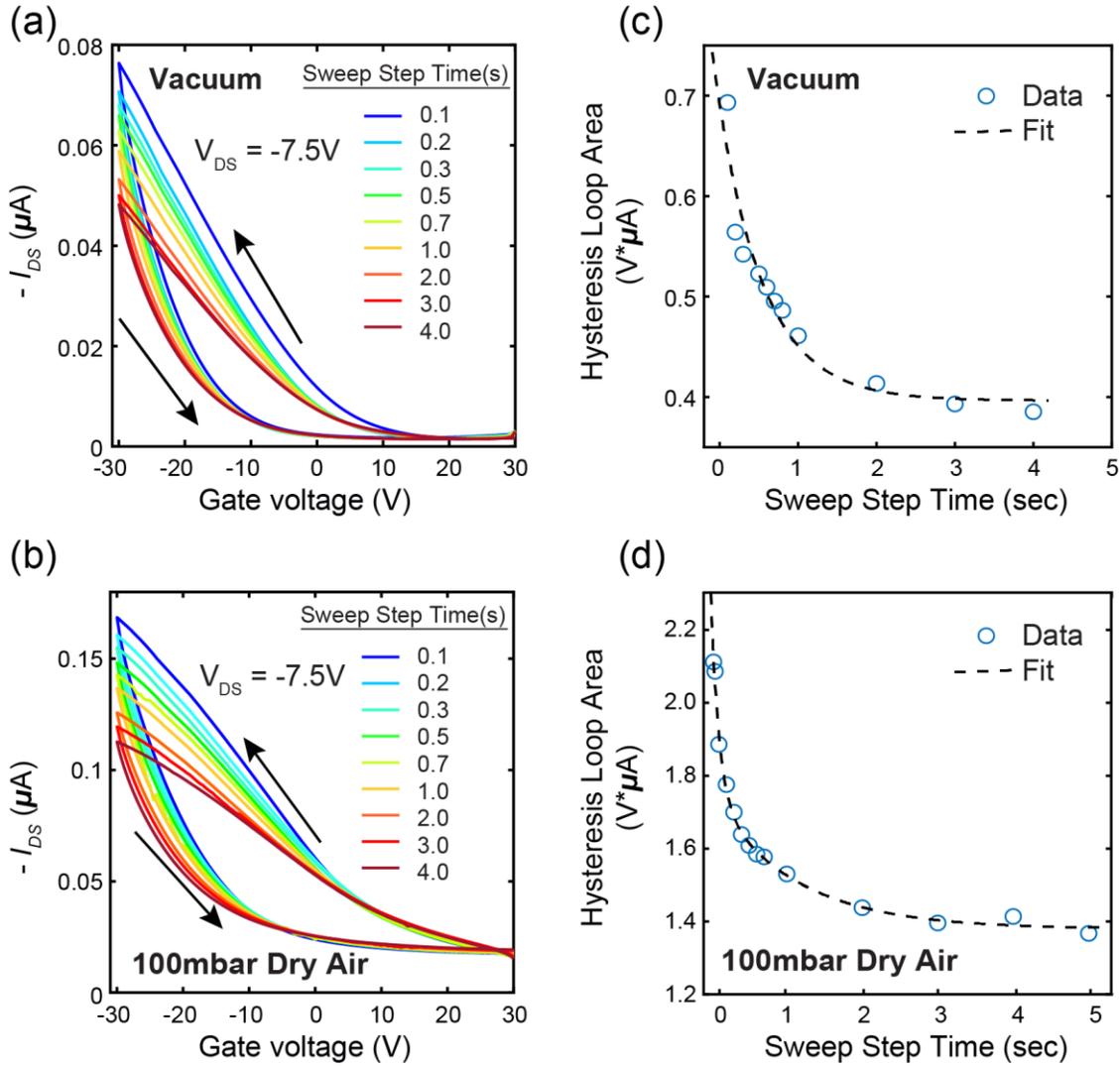

**Figure 6.** Hysteresis analysis of the Cd-doped InAs NCs FETs. Transfer curve loops with varying the sweep step time of Cd-doped InAs NCs FETs measured under vacuum (a), and under 100 mbar of air (b). The corresponding hysteresis loop area of each time step is presented for the device measured under vacuum (c) and under dry air (d).

In the second case, minority carrier (electron) traps close to the semiconductor-oxide interface dominate, and are quickly filled at high positive voltages at the beginning of the forward sweep. The electron trapping boosts $I_{ds}$ since the net hole density thus rises above the amount expected by the given $V_{gs}$. During the forward sweep from positive to negative $V_{gs}$, electron traps are slowly emptied, causing a slow decrease of the net hole density and a nonlinear transfer curve and a $V_{gs}$-dependent mobility. When the gate is swept backwards, electron traps are already fully emptied and the net hole density decreases as well leading to the lower-back-sweep current.[62,64] These characteristics conform with the observations for the Cd-doped NC FET under air. The following picture emerges - upon exposure to air, oxygen molecules that are adsorbed on the surface of the NCs film act as p-type dopants creating acceptor states that trap minority carrier electrons, also leading to the observed nonlinear curve in the forward sweep of $V_{gs}$.

For both vacuum and air conditions, the hysteresis loop area is decreased as the sweep scan time is elongated. An estimation for the carrier trapping rates can be obtained by examining the dependence of the changing hysteresis loop area upon $V_{gs}$ scan rate.[65] Under vacuum (Figure 6c) the loop area decrease is well represented by a single exponential decay with a decay lifetime of $0.6 \pm 0.3$ seconds, consistent with a single factor responsible for the hysteresis in this device related solely to the remaining hole traps upon Cd doping. In air however (Figure 6d), a bi-exponential dependence was observed, with decay lifetimes of $1.1 \pm 0.7$ seconds (within the error limits of the same effect under vacuum), and an additional faster decay component of $0.11 \pm 0.05$ seconds. The faster timescale is assigned to electron-withdrawing acceptor states created by adsorbed oxygen molecules serving as minority (electron) traps. These electron-withdrawing states cause an increase in the hole density, leading to higher hole current.

Altogether, the changes in the conductivity of the Cd-doped device are significantly less pronounced compared to the intrinsic devices, as Cd also serves as a protective layer against oxidation. To confirm this hypothesis, we preformed X-ray photoelectron spectroscopy (XPS), inspecting the As, In, and Cd 3d emission peaks of intrinsic InAs and Cd-doped InAs NCs samples taken from the same batches used for the fabrication of the FETs. The samples were spin-coated on Au coated glass and treated with EDT, in the same manner as in the FET fabrication. Before the measurements, the samples were kept in nitrogen glovebox. Two samples were made for each batch of NCs (intrinsic/Cd doped). One sample of each batch was measured immediately after fabrication while maintaining air-free conditions throughout. The other samples were exposed to ambient atmosphere for 72 hours before measurement. We normalized the elemental composition percentage of all elements to that of arsenic assuming that its concentration does not change during the Cd-doping reaction. Comparing the XPS signals of the samples that were kept under inert conditions shows that for the intrinsic InAs sample, the As:In ratio was 1: 2.2 favorable to indium in line with the ICP-AES measurements (As:In ratio of 1:1.2), and further indicating an In-rich surface owing to the XPS surface sensitivity. For the Cd doped InAs sample, the As:In ratio decreased significantly to 1: 0.7, favorable to arsenic, and the As:Cd ratio signal was 1: 2.2 favorable of cadmium. These results provide further verification that during the doping process, cadmium atoms substitute indium atoms from the surface of the NCs, and a protective shell of cadmium atoms is formed on the surface.

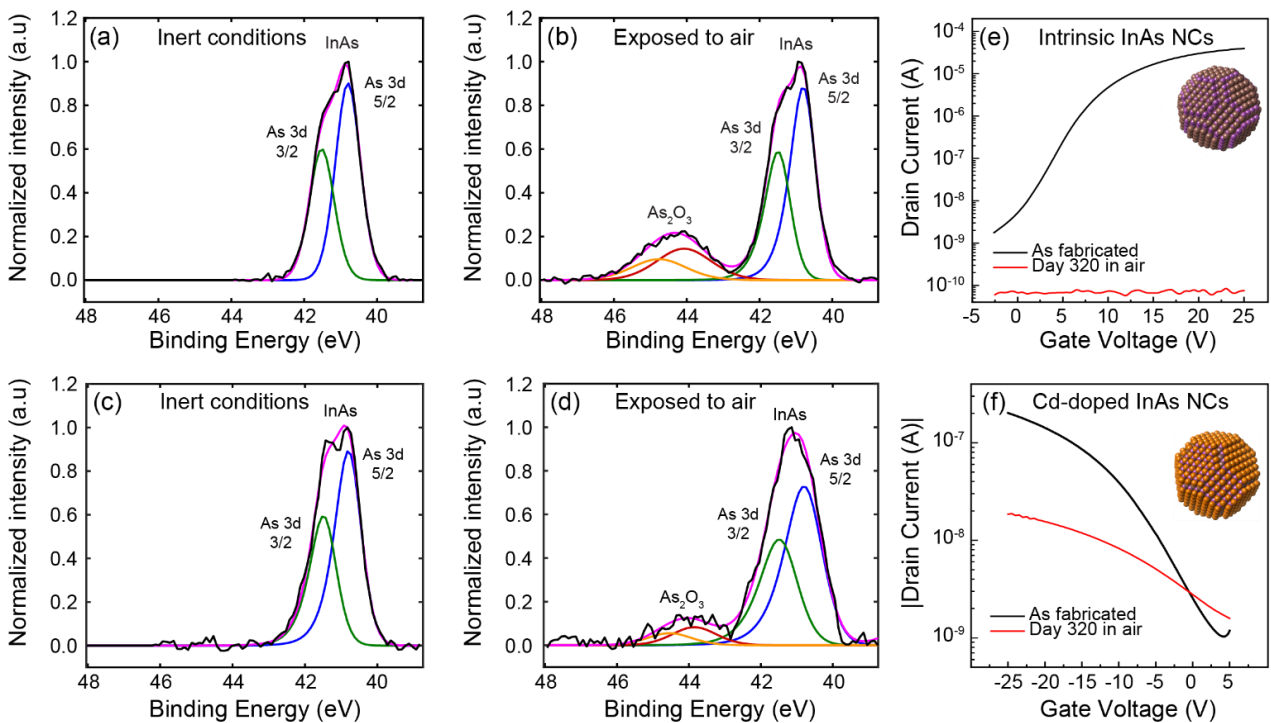

**Figure 7:** Comparison of the oxidation effect of intrinsic and Cd-doped InAs NCs films upon exposure to air. As XPS signal of intrinsic (a, b) and Cd-doped (c,d) InAs NCs films measured under inert atmosphere and after exposure to air. The transfer curves of (e) intrinsic and (f) Cd-doped InAs NCs FETs measured immediately after fabrication (black) and after 320 days while kept in ambient atmosphere (red).

Regarding the air sensitivity, among the elements probed by XPS, arsenic has the most distinct oxidation feature with a peak around 44 eV. Regarding Cd, the XPS data 3d peaks are observed at 405.1 and 411.9eV and did not show a detectable difference for the air exposed samples. Indeed, the $Cd^0$ peaks are located very close to the $Cd^{2+}$ signature, which is also nearly identical for CdO and $Cd_3As_2$,[66] and below the measurement resolution (Figure S9). The In XPS data manifested only slight changes between the inert and air-exposed samples. Therefore, we followed the As signal as a probe for oxidation events on the NCs surfaces. **Figure 7**, presents the As 3d peaks of the intrinsic InAs sample (panels a, b) and Cd-doped InAs sample (panels c, d). Samples that were kept under inert conditions show clear In-As binding energy signals at 41 eV, without any signs of oxidation. After exposure to ambient atmosphere for 72 hours, clear signs of oxidation are evidenced in the intrinsic sample as additional peaks appear, indicating $As_2O_3$ bonds at ~44 eV. Comparing the oxidized As signal between the intrinsic and the Cd doped samples under similar conditions, we find that in the intrinsic sample, $As_2O_3$ signal constitutes 27% of the entire As signal, whereas for the Cd doped sample the $As_2O_3$ signal constitutes only 11% of the As signal. Evidence for decreased oxidation upon exposure to air was also observed for the In peak. The $In_2O_3$ signal of the intrinsic InAs sample that was exposed to air is 3-fold larger compared to the sample that was kept under inert environment, while it remains stable in the Cd-doped samples (see Figure S9). This directly shows that Cd, beyond providing the p-doping, serves also as a protective layer against oxidation of the inner core.

Remarkably, this shielding effect of the Cd layer is also strongly evidenced in the electrical properties of intrinsic InAs and Cd-doped InAs NCs FETs that were kept in ambient condition for 320 days (Figure 7e,f). The current for intrinsic InAs NCs FET under ambient conditions decreased by more than 5 orders of magnitude, while for the Cd-doped InAs FET, the hole current decreased only by an order of magnitude during the same period showing resistance to oxidation.

## 3. Summary and Conclusions

In conclusion, we developed a *post*-synthesis doping reaction to render intrinsically n-type InAs NCs films to p-type. Using a combination of electron microscopy and XAFS methodologies, we decipher the location of the Cd dopant within the InAs NCs and elucidate the doping mechanism. Cd acts as a substitutional dopant, replacing Indium near the surface of the NC, leading to the observed p-type conduction in Cd-doped InAs NCs FETs. Moreover, Cd also acts as a protective layer against oxidation, leading to robust p-type conduction under ambient atmosphere. While the conduction in undoped InAs NCs FETs decreased drastically (both in the gated- and the subthreshold-regime) upon exposure to air even in low pressures, Cd-doped InAs FETs showed remarkable stability with only small changes in conduction. Investigation of the transfer curves sweep in Cd-doped FETs revealed that the hysteresis originates from hole traps. When the device was exposed to dry air, additional contribution to hysteresis is observed, which is related to electron withdrawing states due to adsorbed oxygen molecules. The role of Cd as a protective layer was directly revealed by XPS of films of undoped and Cd-doped InAs NCs upon exposure to ambient conditions showing significant oxidation for the former with much less change for the latter. Indeed, while intrinsic InAs NCs FETs that were kept under ambient conditions for nearly a year degraded completely, the Cd-doped FET still showed p-type characteristics. The post-synthesis doping strategy presented and characterized herein can be relevant to other impurities and for additional semiconductor NC systems. For the technologically important III-V semiconductor NCs, control by doping is of essence for their future incorporation in printed electronics applications.

**Supporting Information**.
Supporting Information is available from the Wiley Online Library or from the author.

**Acknowledgements**
This research was supported by the Israel Science Foundation (Center of Excellence, grant No. 1867/17, U.B.), the Israel Science Foundation (grant no. 488/16, N.T.), and the Technion Ollendorff Minerva Center (N.T.). EXAFS analysis and modeling by A.I.F. were supported as part of the Integrated Mesoscale Architectures for Sustainable Catalysis (IMASC), an Energy Frontier Research Center funded by the U.S. Department of Energy, Office of Science, Basic Energy Sciences under Award #DE-SC0012573. This research used beamline 7-BM (QAS) of the National Synchrotron Light Source II, a U.S. DOE Office of Science User Facility operated for the DOE Office of Science by Brookhaven National Laboratory (BNL) under Contract No. DE-SC0012704. A.I.F and J.L. thank Dr. S. Ehrlich and Dr. L. Ma for help with experiments at the QAS beamline. U.B. and L.A. thank Dr. Vitaly Gutkin and Dr. Sergei Remennik from the Unit for Nanocharacterization of the Center for Nanoscience and Nanotechnology at the Hebrew University of Jerusalem for assistance in the materials characterization and for helpful discussions. We also acknowledge the technical support of the staff of the Unit for Nanofabrication at the Center for Nanoscience and Nanotechnology at the Hebrew University of Jerusalem. U.B. thanks the Alfred & Erica Larisch memorial chair.

**Conflict of interests**

The authors declare no conflict of interests

# Supporting Information

# InAs nanocrystals with robust p-type doping

*Lior Asor, Jing Liu, Yonatan Ossia, Durgesh C. Tripathi, Nir Tessler,*

*Anatoly I. Frenkel, and Uri Banin\**

# Table of contents

# 1. Experimental section

**1.1 Materials.** Toluene (anhydrous, 99.8%), hexane (anhydrous, 99.8%), Methanol (anhydrous, 99.8%), acetonitrile (anhydrous, 99.8%), (3-mercaptopropyl)trimethoxysilane (95%), 1,2-Ethanedithiol (98%), Indium(III) chloride ($InCl_3$, 99.999%), Cadmium(II) Oxide (CdO, 99.99%), Oleic acid (90%), 1-octadecene (ODE, 90%), Oleylamine (OlAm, 98%) were purchased from Sigma-Aldrich (Merck) and used as received. Tri-n-octylphosphine (TOP, 90%) was purchased from Sigma-Aldrich (Merck) and was distilled prior to use. Tris(trimethylsylil)arsine ($(TMS)_3As$) was synthesized and distilled in our lab by using a well-established procedure.

**1.2 Synthesis of Colloidal InAs NCs.** Colloidal InAs NCs were synthesized following a well-established wet-chemical synthesis.[1] Precursor solutions (nucleation and growth solutions) containing mixtures of (tris(trimethylsilyl)arsine) and $InCl_3$ dissolved in distilled TOP were prepared in advanced and kept under inert atmosphere. At the Schlenk line, a solution of distilled TOP was evacuated at room temperature under high vacuum for 30 min and then heated to 300 °C. Once the temperature reached 300°C, the nucleation solution containing a mixture of $InCl_3$ in TOP and $(TMS)_3As$ (molar ratio 2:1 In:As) was rapidly injected. The heating mantle was removed, and the temperature was lowered to 260 °C. At 260°C, the growth solution (1.2:1 In:As) was slowly added, allowing for particle growth until the desired size was reached by monitoring the gradual red-shift of the 1st exciton peak. Narrow size distributions were achieved by size selective precipitation performed in a glovebox by addition of methanol to the NC dispersion and filtering the solution through a 0.2 μm polyamide membrane filter (Whatman). The precipitant was subsequently dissolved in anhydrous toluene (Sigma) and kept under constant inert conditions throughout.

**1.3 Post-synthesis doping of InAs NCs.** Inspired by a method for shell growth on InAs NCs,[2] On the Schlenk line, 4 mL of ODE and 1.5 mL of OlAm were added to a 50 mL round bottom flask and were heated under vacuum to 100°C for 1 hr. 100 nmol of highly monodispersed fraction of as synthesized InAs NCs in toluene was added to the flask and the solution was evacuated at room temperature for 30 min. and then the temperature was raised again to 100 °C, still under vacuum for additional 30 minutes in order to remove any of the toluene residues. Then, the temperature was raised to 260°C under Ar flow and once the temperature settled, calculated amount of Cd(Oleate)$_2$ in ODE (0.15M), correlating to In:Cd ratio of 1:2 were added dropwise over a period of 20 minutes. The system was left for 3 hrs. during this time, aliquots were taken and the absorbance and photoluminescence of the doped NCs were monitored. Purification of the doped NCs from the crude solution was performed in the glovebox. To the crude solution, 3 mL of toluene were added, and the solution was filtered. The filtered solution was transferred to a 500 mL round flask and ethanol was added dropwise until the solution turned murky and was filtered again. The precipitate left on the filter was redispersed in hexane. The remaining filtered solution was transferred back to the large flask and the precipitation-filtration process was repeated again, until the crude solution turned colorless. The different fractions of filtered Cd-doped InAs NCs were kept in different vials for further characterization and device fabrication.

**1.4 FET Fabrication.** Heavily doped p-type Si substrates covered with 100 nm thick SiO$_2$ and 10 nm HfO$_2$ coating were cleaned by subsequently dipping and sonicating the substrate in acetone, methanol and isopropanol for 5 min. After this cleaning procedure, the substrates were blown dry with nitrogen and put inside a UV-Ozone cleaner for 30 min. The cleaned substrate were then transferred into the N$_2$ glovebox, where they were soaked in a 5 mM (3-mercaptopropyl)trimetoxysilane in methanol, and kept in this solution overnight. Thin films of InAs NCs on top of Si/SiO2 substrates were prepared as follows: 15 mg/mL solution of InAs NCs in hexane were prepared and filtered twice using a 0.1 µm diameter porous PTFE filter to remove large aggregates and impurities. Clean InAs NCs solution was added dropwise (~30 µL) on the Si/SiO2 substrate and span at 2000 rpm for 30 s. Solid-state ligand exchange with 1,2-Ethanedithiol (EDT) was performed by covering the film with 5 mM solution of EDT in acetonitrile for 30 s and spinning at 2000 rpm for 15 s to dry the film. The film was washed twice with pure acetonitrile and spinning the substrate at 2000 rpm for another 30 s. The entire cycle was repeated until the desired thickness (30−35 nm) of InAs NCs was achieved. Thermal annealing of the as prepared films was performed prior to metal contacts evaporation. As-prepared films of InAs NCs were put inside an oven inside the glovebox and heated to 220 °C for 30 min under a $N_2$ environment. During this process, organic residues are removed from the film and the film hardens. Au source/drain electrodes (120 nm thick, interdigitated configuration; W = 20 mm, L = 100 μm) were thermally evaporated using Electron Beam evaporation of Au through a specially designed evaporation mask

## **2. Characterization**

**2.1 ICP−AES Measurements.** ICP−AES measurements were carried out using a PerkinElmer Optima 3000 to determine dopant concentrations within the NCs. Samples were prepared by dissolving the doped NCs in HNO3 and diluting the solution with triply distilled water.

**2.2 XAFS Measurements.** These were performed at the Brookhaven National Laboratory National Synchrotron Light Source II beamline 7-BM. The specimens were prepared under inert conditions in a $N_2$ glovebox, where they were spread on either Kapton tape for the native Cd-doped sample or on thin glass slits for the EDT exchanged Cd-doped NCs films, sealed and mounted on a sample. The holder was put inside a bag, filled with nitrogen to maintain inert conditions through the experiments. Experiments at the As and In K-edges were performed in transmission mode, and for the Cd K-edge, both in transmission and fluorescence. 30 consecutive measurements were taken at each edge to improve the signal-to-noise ratio in the data. XAFS data processing and analysis were performed using standard techniques. Briefly, the data were aligned in energy using reference spectra collected in standard materials simultaneously with the samples' data, and then averaged. Alignment, merging and background subtraction was done using the Athena program.[3] All fittings to the acquired data were done using Artemis program.[3,4] The fitting process was performed simultaneously for different absorption edges by applying appropriate constraints between the fitting parameters. For Cd K-edge EXAFS fits, Cd-As, and Cd-O paths were included in the fitting model for the as-prepared Cd-doped InAs NCs powders. For the model of the EDT ligand exchanged films, additional Cd-S path was included. The amplitude reduction factor ($S_0^2$ = 0.82) was found from the fit to Cd(Oleate)$_2$ standard sample and fixed in the fits to the doped NC data. Variables in the theoretical EXAFS signal included the coordination numbers of nearest neighbors of a certain type around the absorbing Cd atom (N), the bond distance between the absorber and the nearest neighbors (R), and the mean-square-displacement ($\sigma^2$). $S_0^2$ values were fixed for the same absorption element (Cd, In or As) in all samples while the other parameters were varied. For In and As K-edge EXAFS fits, only one path was used for fitting (In-As and As-In, respectively) and their coordination numbers were constrained to be 4 for the undoped sample in order to acquire the $S_0^2$ factor (0.92 for both In and As ). Similarly to Cd edge data, for the doped samples, N, R and σ2 were set as variables, while $S_0^2$ was fixed for all samples. The fitting range in k space is 2-12 Å$^{-1}$ for Cd, 2-13 Å$^{-1}$ for In and As. The fitting range in R space is 1.45-3.2 Å for Cd, 1.65-3 Å for In, and 1.45-3 Å for As.

## 3. Figures and Tables

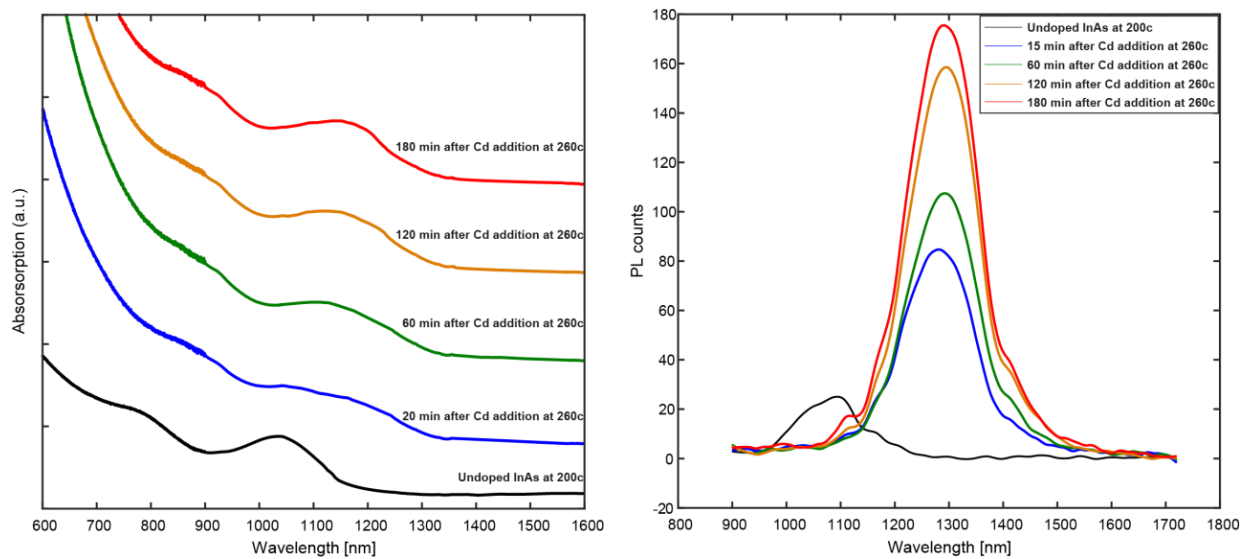

**Figure S1:** Absorption (a) and photoluminescence (b) spectra of InAs NCs during the Cd doping reaction. Aliquots taken at different times during the reaction. It is observed that the 1$^{st}$ exciton absorption peak and photoluminescence (PL) peak are red shifted during the reaction. The PL quantum yield is also improved.

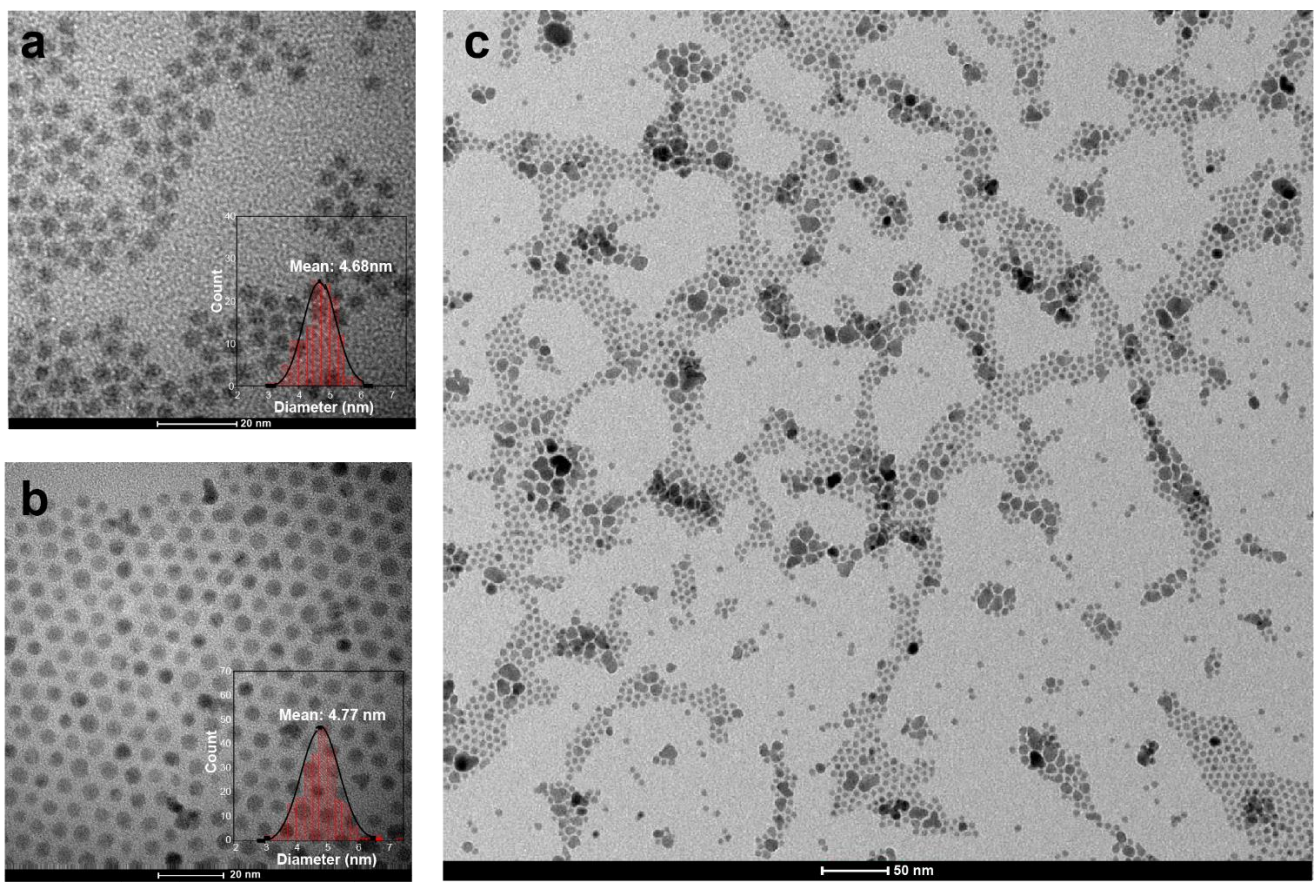

**Figure S2:** TEM images of: (a) the as-synthesized InAs NCs, (b) purified fraction of the Cd-doped InAs NCs. (c) Cd-doped InAs sample immediately after the doping reaction ends, including large aggregates that formed during the reaction.

**Table S1.** Calculated ratios of As:In and As:Cd acquired by ICP-AES

| Sample name | In | As | Cd |
|---|---|---|---|
| As-synthesized InAs NCs | 1.2 | 1.0 | -- |
| Cd-doped InAs NCs Low doping | 1.15 | 1.00 | 0.28 |
| Cd-doped InAs NCs Medium doping | 1.1 | 1.0 | 0.4 |
| Cd-doped InAs NCs High doping concentration | 1.07 | 1.00 | 0.6 |

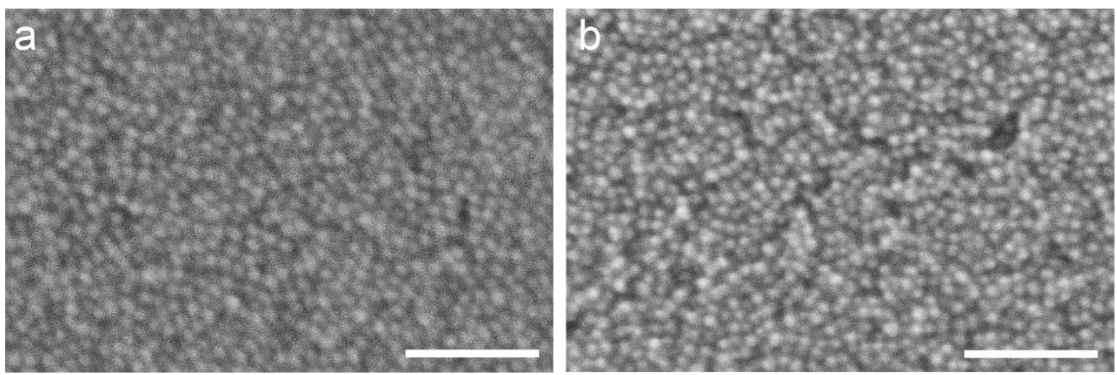

**Figure S3:** High-resolution SEM image of the NCs spin coated on top of Si/SiO$_2$ substrates, (a) before and (b) after thermal annealing at 220 °C for 30 min under a N2 atmosphere. The scale bar of both images is 50 nm.

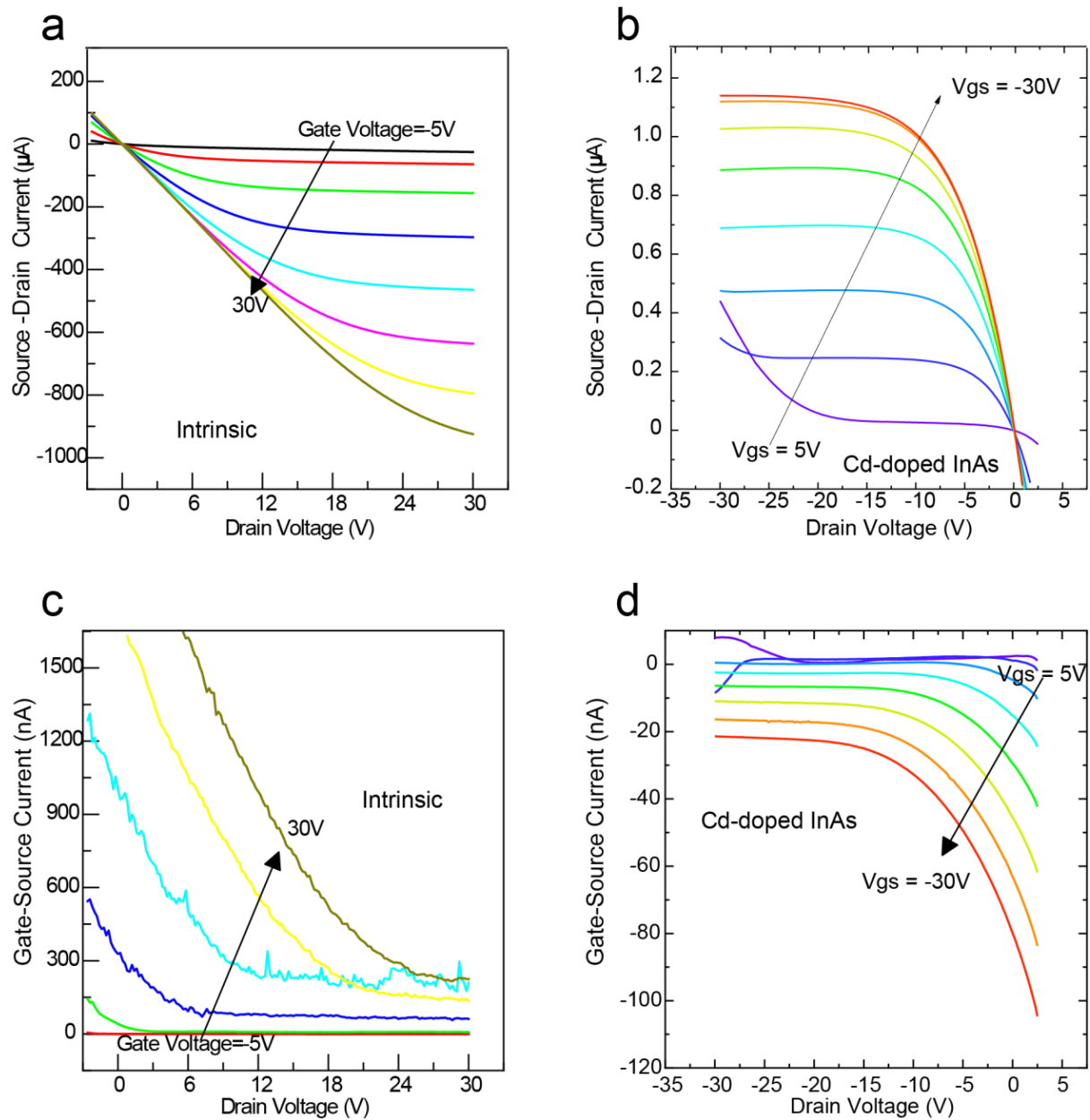

**Figure S4:** Measured Source-Drain (a-b) and Gate-Source (c-d) currents vs. Drain voltage of Intrinsic and Cd-doped InAs NCs FETs presented in Figure 2 of the main text, respectively.

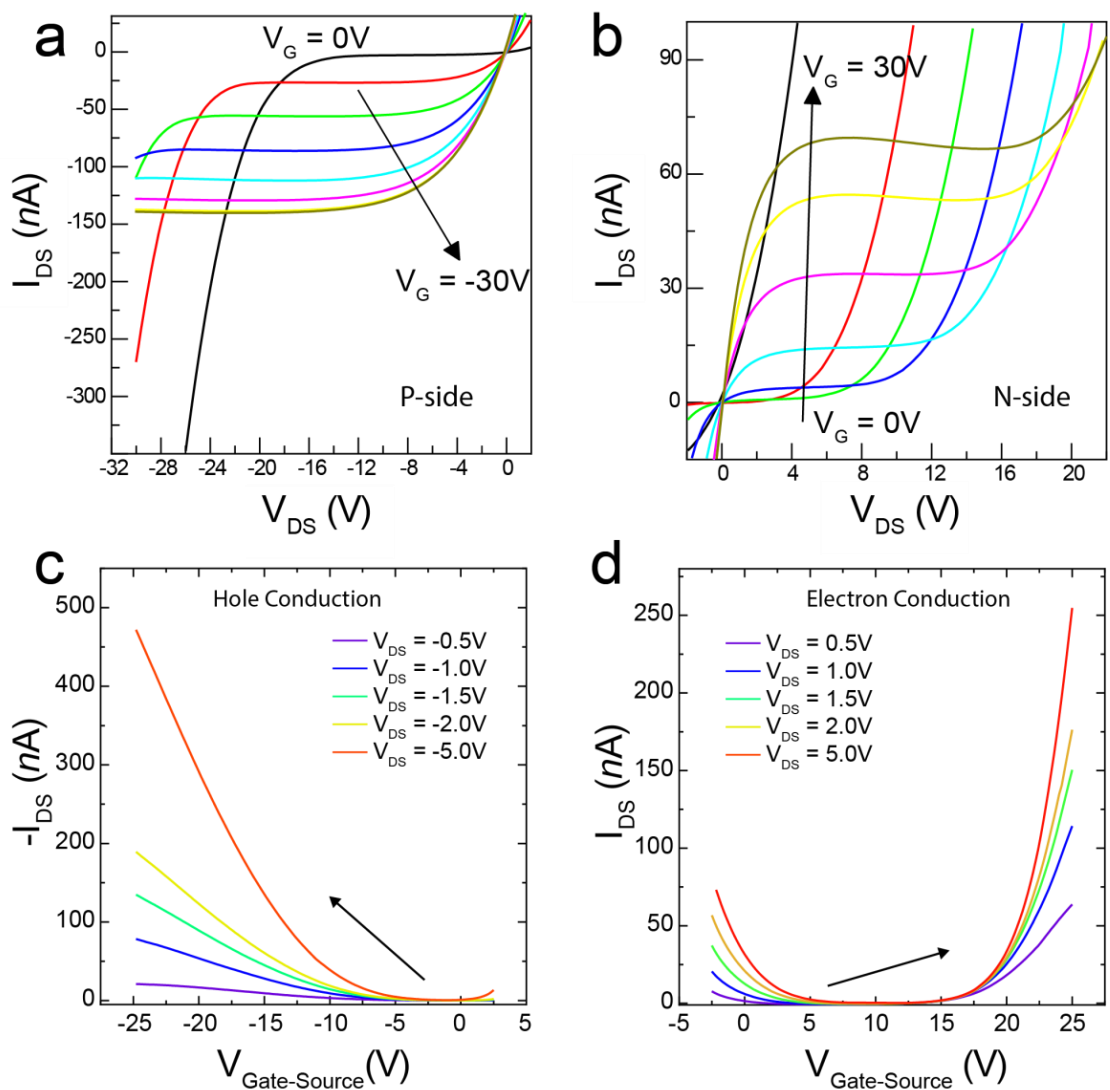

**Figure S5:** (a, b) Output characteristics of the p-side and the n-side of lightly Cd-doped InAs NCs FETs, respectively, showing ambipolar conduction. Panels c, d show the transfer characteristics of the same device indicating hole conduction with hole-threshold voltage = -11V and electron conduction with electron-threshold voltage = +18V. Black arrows indicate the gate-sweep direction.

[S11]

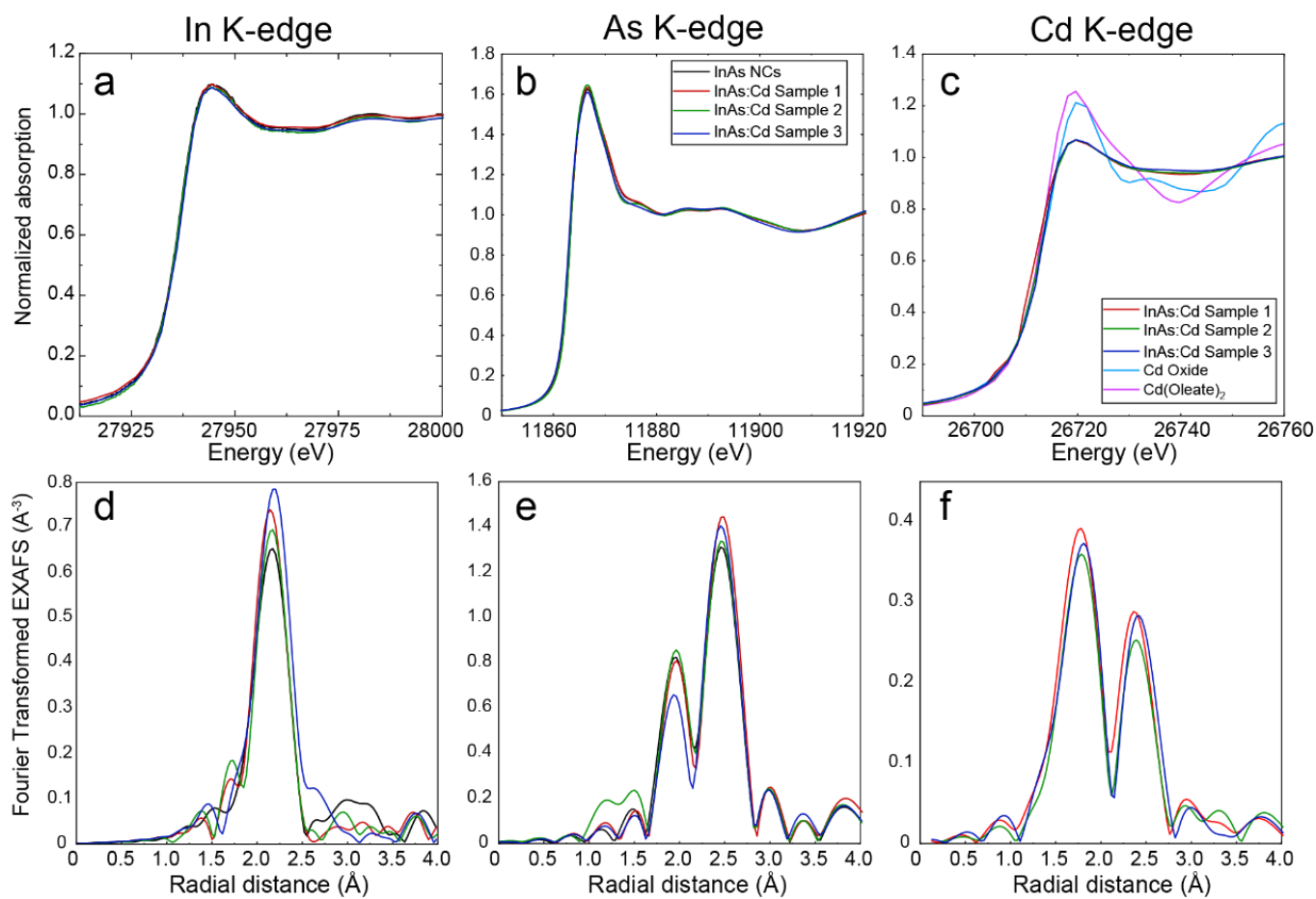

**Figure S6:** XANES spectra of In, As and Cd k-edge (a-c) for as-synthesized InAs NCs and different samples of Cd-doped samples. Fourier transform magnitudes of the EXAFS spectra for the respective elements are given in panels d-f.

**Table S2.** Best fit values of the fitting parameters of In and As and Cd edges from as-prepared Cd-doped InAs NCs samples kept under inert conditions. Asterisk denotes fixed parameters.

| Sample name | Edge | Bond | C.N. | R (Å) | σ²(Å²) | ΔE (eV) | reduced chi² | R-factor |
|---|---|---|---|---|---|---|---|---|
| Undoped InAs | In | In-As | 4* | 2.615±0.002 | 0.0051±0.0004 | -0.7±0.5 | 235.07 | 0.015 |
| | As | As-In | 4* | 2.615±0.002 | 0.0032±0.0006 | 3.8±0.3 | | |
| Cd doped Not annealed | In | In-As | 4.1±0.5 | 2.613±0.003 | 0.0041±0.0006 | -0.7±0.5 | 235.07 | 0.015 |
| | As | As-In | 3.9±0.5 | 2.613±0.003 | 0.0042±0.0006 | 3.8±0.3 | | |
| | Cd | Cd-As | 1.4±1.2 | 2.62±0.01 | 0.008±0.008 | 9.4±1.0 | 232.07 | 0.034 |
| | | Cd-O | 1.2±0.4 | 2.30±0.02 | 0.001±0.005 | | | |
| Cd doped Vacuum annealed | In | In-As | 4.0±0.4 | 2.616±0.008 | 0.0043±0.0003 | 0±2 | 375.3 | 0.012 |
| | As | As-In | 4.5±0.3 | 2.605±0.006 | 0.0042±0.0005 | 3.5±0.5 | | |
| | Cd | Cd-As | 0.9±0.5 | 2.62±0.01 | 0.008±0.008 | 9.4±1.0 | 232.07 | 0.034 |
| | | Cd-O | 1.5±0.4 | 2.30±0.02 | 0.004±0.005 | | | |
| Cd doped N$_2$ annealed | In | In-As | 4.0±0.5 | 2.62±0.01 | 0.0043±0.0004 | 0±2 | 333.9 | 0.016 |
| | As | As-In | 4.2±0.1 | 2.606±0.006 | 0.0045±0.0009 | 3.7±0.8 | | |
| | Cd | Cd-As | 1.0±0.7 | 2.62±0.01 | 0.006±0.006 | 9.4±1.0 | 232.07 | 0.034 |
| | | Cd-O | 1.2±0.3 | 2.30±0.01 | 0.001±0.004 | | | |

**Table S3.** Best fit values of the fitting parameters of In and As and Cd edges from EDT-exchanged NCs films kept under inert conditions. Asterisk denotes fixed parameters.

| Sample name | Edge | Bond | C.N. | R (Å) | σ²(Å²) | ΔE (eV) | reduced chi² | R-factor |
|---|---|---|---|---|---|---|---|---|
| Undoped InAs | In | In-As | 4* | 2.615±0.003 | 0.0051±0.0004 | 4.8±0.8 | 21.17 | 0.013 |
| | As | As-In | 4* | 2.615±0.003 | 0.0040±0.0003 | 3.7±0.6 | | |
| Cd doped not annealed | In | In-As | 4±0.5 | 2.61±0.02 | 0.004±0.002 | 5±4 | 50.7 | 0.016 |
| | As | As-In | 3.6±0.1 | 2.611±0.003 | 0.0041±0.0003 | 3.8±0.5 | | |
| | Cd | Cd-As | 1±1 | 2.58±0.05 | 0.01±0.01 | 1±2 | 49.4 | 0.03 |
| | | Cd-S | 3±1 | 2.53±0.03 | 0.01±0.01 | | | |
| Cd doped vacuum annealed | In | In-As | 4±0.5 | 2.61±0.02 | 0.005±0.002 | 3±2 | 92.8 | 0.018 |
| | As | As-In | 3.4±0.1 | 2.609±0.003 | 0.0040±0.0004 | 3.8±0.5 | | |
| | Cd | Cd-As | 1.0±0.8 | 2.59±0.07 | 0.006±0.008 | 1±1 | 12.7 | 0.018 |
| | | Cd-S | 2±1 | 2.54±0.03 | 0.01±0.01 | | | |
| Cd doped N$_2$ annealed | In | In-As | 4±0.5 | 2.60±0.02 | 0.004±0.002 | 3±5 | 112.5 | 0.012 |
| | As | As-In | 3.3±0.1 | 2.609±0.002 | 0.0038±0.0003 | 3.7±0.4 | | |
| | Cd | Cd-As | 1.3±0.9 | 2.58±0.04 | 0.005±0.006 | 1±1 | 23.9 | 0.005 |
| | | Cd-S | 2.5±0.8 | 2.53±0.01 | 0.01±0.01 | | | |

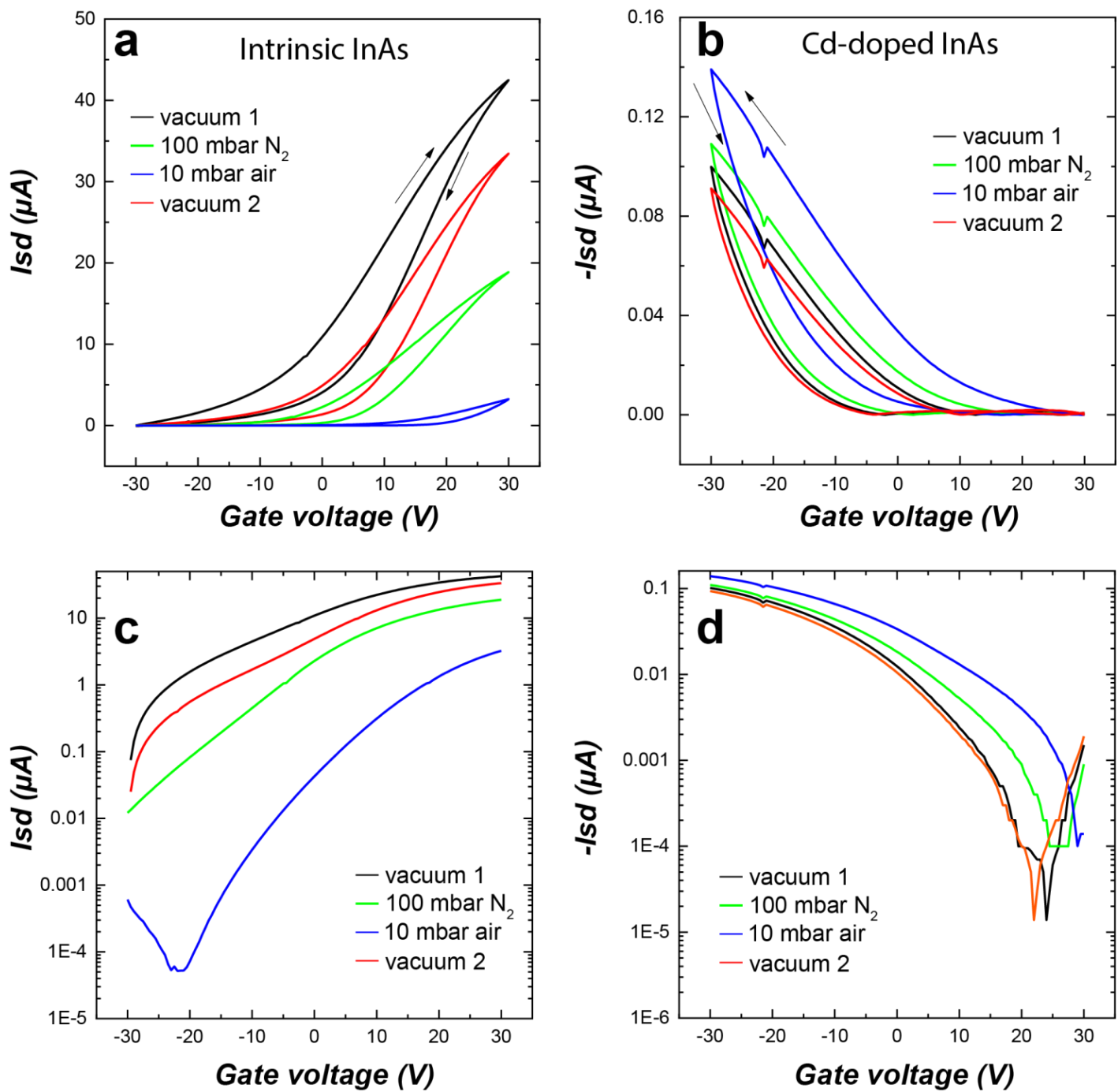

**Figure S7:** complete transfer sweep characteristics of intrinsic (a) and Cd-doped (b) InAs NCs FETs measured under different atmospheres. The logarithmic scale of panels a and b are presented in panels c-d.

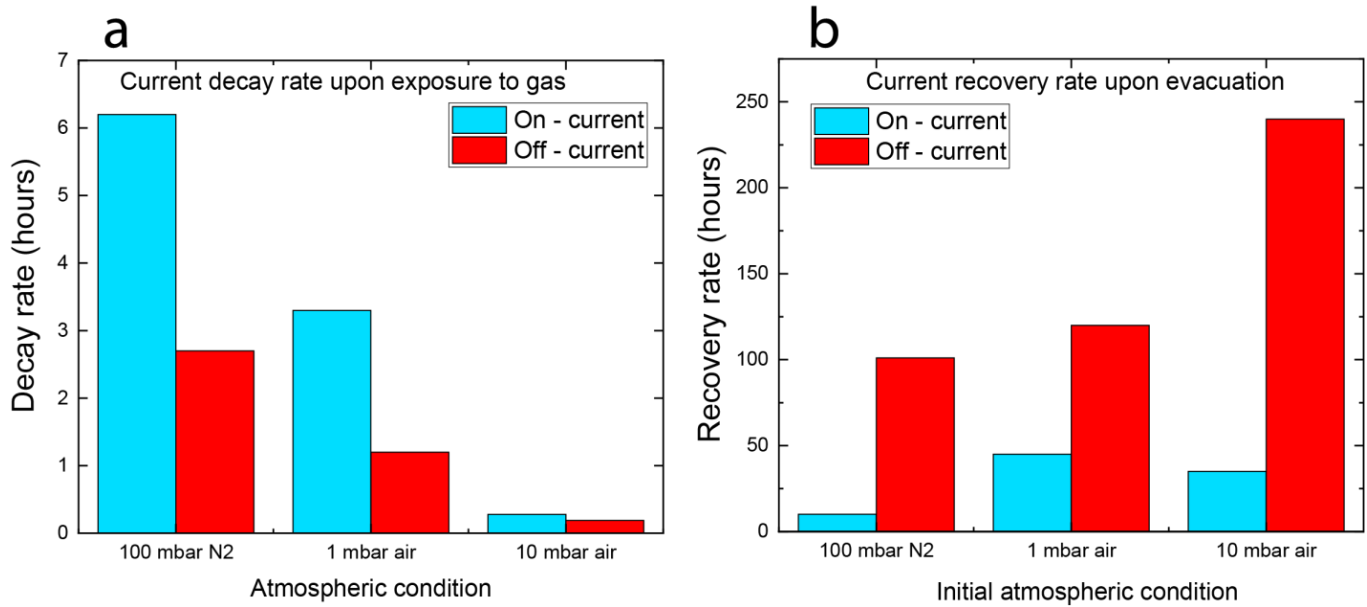

**Figure S8:** Lifetime analysis of decay (exposed to different gas pressures, panel a) and recovery (under $10^{-4}$ mbar vacuum after exposure, panel b) rates of the 'On' (cyan bars) and 'Off' (red bars) currents of intrinsic InAs NC based FET.

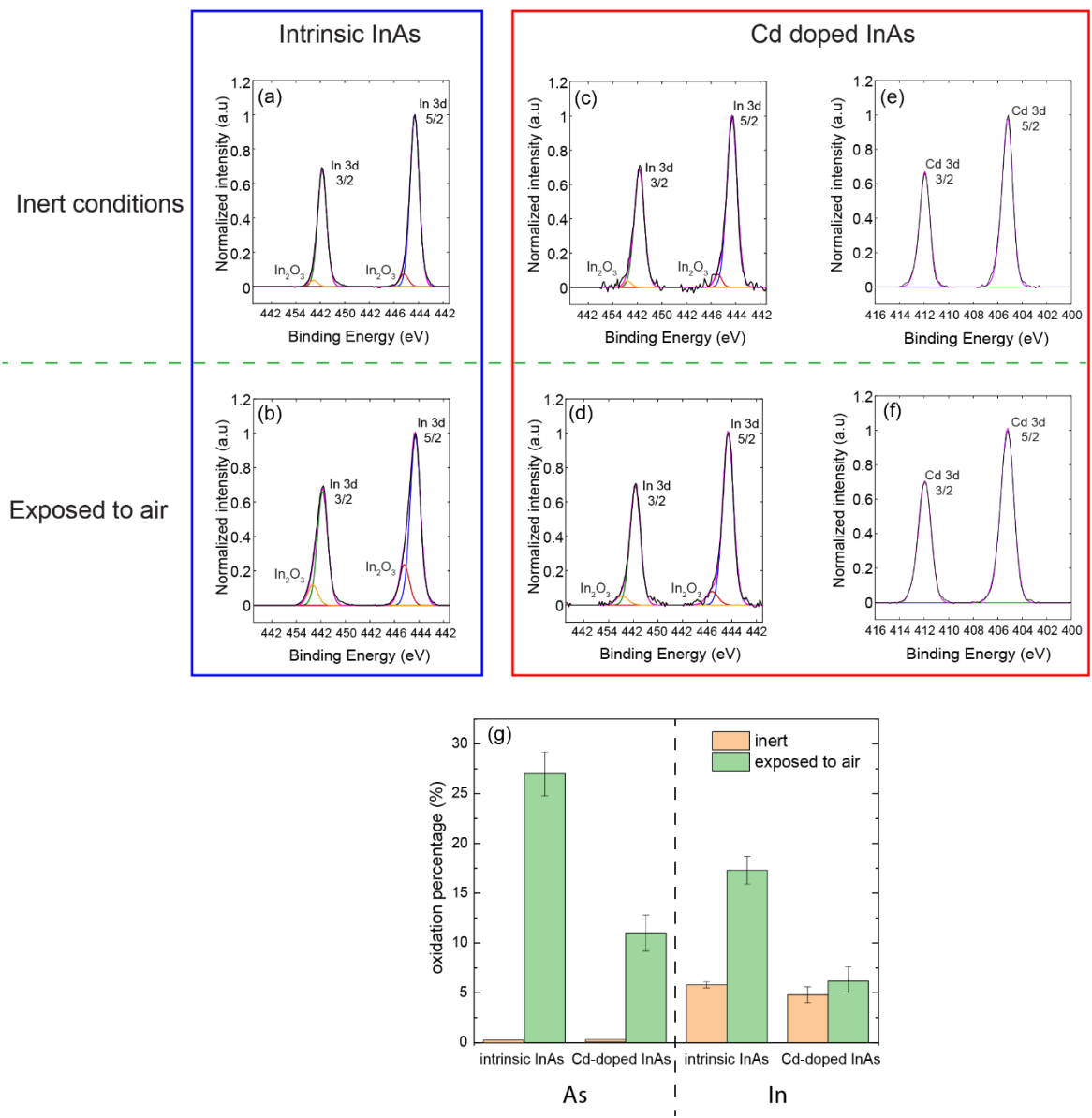

**Figure S9:** XPS analysis of In and Cd in intrinsic and Cd-doped InAs NCs samples. Panels a and b presents the In signal of intrinsic InAs sample measured under inert conditions (a) and after exposure to ambient air for 72 hours (b). panels c,d and e,f present the In and Cd XPS signals of Cd doped InAs NCs films, respectively, measured under inert conditions and after exposure to air. Panel g presents a comparison of the oxidation percentage of In and As before and after exposure to air.

[S16]